\documentclass[12pt]{article}

\usepackage{graphicx}

\def\gtwid{\mathrel{\raise.3ex\hbox{$>$\kern-.75em\lower1ex\hbox{$\sim$}}}}
\def\ltwid{\mathrel{\raise.3ex\hbox{$<$\kern-.75em\lower1ex\hbox{$\sim$}}}}
\def\square{\kern1pt\vbox{\hrule height 1.2pt\hbox{\vrule width 1.2pt\hskip 3pt
   \vbox{\vskip 6pt}\hskip 3pt\vrule width 0.6pt}\hrule height 0.6pt}\kern1pt}

\begin{document}

\begin{titlepage}

\begin{flushright}
UFIFT-QG-17-01 \\
CCTP-2017-1
\end{flushright}

\vskip 1cm

\begin{center}
{\bf Invariant Measure of the One Loop Quantum Gravitational Back-Reaction on Inflation}
\end{center}

\vskip .5cm

\begin{center}
S. P. Miao$^{1*}$, N. C. Tsamis$^{2\star}$ and R. P. Woodard$^{3\dagger}$
\end{center}

\vskip .5cm

\begin{center}
\it{$^{1}$ Department of Physics, National Cheng Kung University \\
No. 1, University Road, Tainan City 70101, TAIWAN}
\end{center}

\begin{center}
\it{$^{2}$ Institute of Theoretical Physics \& Computational Physics, \\
Department of Physics, University of Crete, \\
GR-710 03 Heraklion, HELLAS}
\end{center}

\begin{center}
\it{$^{3}$ Department of Physics, University of Florida,\\
Gainesville, FL 32611, UNITED STATES}
\end{center}

\vspace{.5cm}

\begin{center}
ABSTRACT
\end{center}

We use dimensional regularization in pure quantum gravity on de Sitter
background to evaluate the one loop expectation value of an invariant 
operator which gives the local expansion rate. We show that the 
renormalization of this nonlocal composite operator can be accomplished
using the counterterms of a simple local theory of gravity plus matter, 
at least at one loop order. This renormalization completely absorbs the 
one loop correction, which accords with the prediction that the lowest 
secular back-reaction should be a 2-loop effect. 

\begin{flushleft}
\end{flushleft}

\vskip .5cm

\begin{flushleft}
$^{*}$ e-mail: spmiao5@mail.ncku.edu.tw \\
$^{\star}$ e-mail: tsamis@physics.uoc.gr \\
$^{\dagger}$ e-mail: woodard@phys.ufl.edu
\end{flushleft}

\end{titlepage}

\section{Introduction} \label{intro}

The quantum gravitational back-reaction on accelerated expansion has a
special importance because of its potential to simultaneously provide a 
resolution for the (old) problem of the cosmological constant 
\cite{Weinberg:1988cp,Carroll:2000fy} and a predictive model for 
primordial inflation \cite{Mukhanov:2005sc}. The idea is easy to sketch.
We posit that the bare cosmological constant is not absurdly small, but 
rather large and positive, and that this triggered primordial inflation 
\cite{Tsamis:1996qq,Tsamis:2011ep}. Accelerated expansion rips virtual 
scalars and gravitons out of the vacuum \cite{Parker:1969au,Grishchuk:1974ny};
this is what causes the primordial power spectra \cite{Starobinsky:1979ty,
Mukhanov:1981xt}. The self-gravitation of these particles must tend to 
slow the expansion rate, and their contribution to the vacuum energy
must grow with time as more and more of them come into causal contact
through the continual increase in the volume of the past light-cone.
$\Lambda$-driven inflation is based on the assumption that this effect
eventually stops inflation \cite{Tsamis:2011ep}.

Quantum instabilities of de Sitter have been proposed for decades
\cite{Polyakov:1982ug,Myhrvold:1983hx,Ford:1984hs,Mottola:1984ar,
Antoniadis:1985pj,Mazur:1986et,Antoniadis:1991fa,Tsamis:1992sx,
Krotov:2010ma,Akhmedov:2012pa,Polyakov:2012uc,Anderson:2013zia}. The 
difficult part has been to properly quantify the effect so as to establish 
its reality. In 1996 a fixed-gauge computation by Mukhanov, Abramo and 
Brandenberger seemed to show secular slowing at one loop in scalar-driven 
inflation \cite{Mukhanov:1996ak,Abramo:1997hu}. However, Unruh correctly 
questioned the validity of treating the expectation value of the 
gauge-fixed metric as one would a classical metric \cite{Unruh:1998ic}. 
Although the result persisted in a different gauge \cite{Abramo:1998hi,
Abramo:1998hj}, the introduction of a truly invariant measure for the 
local expansion rate, with the time fixed by the value of the inflaton 
\cite{Abramo:2001db}, revealed the absence of any secular slowing 
\cite{Abramo:2001dc,Geshnizjani:2002wp}. The apparent effect in a fixed 
gauge ``time'' arose from quantum fluctuations tending to push the 
inflaton down its potential a little faster than it would have gone 
classically. This is apparent from using different clocks 
\cite{Geshnizjani:2003cn,Marozzi:2011zb,Marozzi:2012tp}, but the inflaton 
potential suffers no secular corrections at one loop order.

True quantum gravitational back-reaction is predicted to occur at two 
loop order because inflationary particle production is a one loop effect
so the quantum gravitational response to it must occur one loop higher
\cite{Tsamis:1996qq,Tsamis:2011ep}. Although the reasoning is solid, 
the conclusion is frustrating because two loop computations in nontrivial
geometries are so difficult. There seems to be no advantage to working in
scalar-driven inflation; that would only have paid off if the scalar-metric
mixing had permitted a one loop effect. In the absence of a reduction in 
the loop order, the presence of a scalar inflaton merely complicates the
problem through the evolving background and the more complex propagators
and vertices. The simplest venue is therefore pure gravity on de Sitter 
background, {\it provided} a suitable invariant expansion observable can
be constructed. A proposal for this has been made based on using a nonlocal
scalar functional of the metric in the same way one would quantify the
expansion rate using a scalar inflaton \cite{Tsamis:2013cka}. However,
this observable can only be used at two loop order if it can be successfully 
renormalized at one loop order. That is the purpose of this paper.

In section~\ref{expop} we review how the expansion operator is defined,
and we give its expansion to second order in metric perturbations about de
Sitter background. Section~\ref{1loop} evaluates the expectation value of
the observable at one loop order using dimensional regularization. Its
renormalization is accomplished in section~\ref{renorm}. Section~\ref{epi}
discusses our results and the prospects for pushing on to two loop order.

\section{The Expansion Observable} \label{expop}

The purpose of this section is to precisely define the expansion operator
and give its expansion in powers of the graviton field \cite{Tsamis:2013cka}.
We also present the local gravity + matter theory from which it descends. 
To facilitate the application of dimensional regularization we work in $D$ 
spacetime dimensions with a single time coordinate $-\infty < \eta < 0$, a 
$(D-1)$-dimensional space vector $\vec{x}$, and a spacelike metric.

Our nonlocal scalar $\Phi[g](x)$ is constructed to obey the equation,
\begin{equation}
\square \Phi \equiv \frac1{\sqrt{-g}} \, \partial_{\mu} \Bigl[ \sqrt{-g} \,
g^{\mu\nu} \partial_{\nu} \Phi \Bigr] = (D \!-\! 1) H \; , \label{Phieqn}
\end{equation}
subject to the initial conditions (at $\eta = \eta_i \equiv -1/H$),
\begin{equation}
\Phi(\eta_i,\vec{x}) = 0 \qquad , \qquad -g^{\alpha\beta}(\eta_i,\vec{x}) 
\partial_{\alpha} \Phi(\eta_i,\vec{x}) \partial_{\beta} \Phi(\eta_i,\vec{x}) 
= 1 \; . \label{IC}
\end{equation}
The important thing about (minus) $\Phi[g](x)$ is that it grows in the 
timelike direction, not just for de Sitter but for an arbitrary metric. Hence 
its gradient produces a timelike 4-vector. By normalizing this vector and 
then taking the divergence we can construct a scalar measure of the local
expansion rate, just as is done with the inflaton in scalar-driven 
inflation \cite{Geshnizjani:2002wp},
\begin{equation}
\mathcal{H}[g](x) = \frac{1}{(D \!-\! 1) \sqrt{-g(x)}} \, \partial_{\mu} 
\Biggl[ \frac{\sqrt{-g(x)} \, g^{\mu\nu}(x) \partial_{\nu} \Phi[g](x)}{
\sqrt{-g^{\alpha\beta}(x) \partial_{\alpha} \Phi[g](x) \partial_{\beta} 
\Phi[g](x)}}  \Biggr] \; . \label{scalarH}
\end{equation}

Expression (\ref{scalarH}) is a scalar, not an invariant, because the 
observation point $x^{\mu} = (\eta,\vec{x})$ has not been invariantly 
fixed. Recall that doing this was the key step in demonstrating that 
there is no one loop secular back-reaction in scalar-driven inflation \cite{Abramo:2001dc,Geshnizjani:2002wp}. Just like the case of scalar-driven 
inflation, we use the value of the scalar to invariantly fix the surface of 
simultaneity on which the observation is made. Also like scalar-driven 
inflation, a homogeneous and isotropic state provides no reference structure 
with which we can fix the spatial coordinates on this surface. So we only
define a metric-dependent time $\theta[g](x)$ by the condition that it makes
the full scalar agree with its value on de Sitter background ($g_{\mu\nu} =
a^2 \eta_{\mu\nu}$, with $a(\eta) = -1/H\eta$),
\begin{equation}
\Phi[g]\Bigl( \theta[g](x),\vec{x}\Bigr) \equiv \Phi_0(\eta) \equiv
-\frac1{H} \, \ln\Bigl[ a(\eta)\Bigr] \; . \label{time}
\end{equation}
(Making the full inflaton agree with its background value was also the time
condition for scalar-driven inflation \cite{Abramo:2001dc,Geshnizjani:2002wp}.)
Evaluating the scalar (\ref{scalarH}) at this time defines the expansion
observable \cite{Tsamis:2013cka},
\begin{equation}
\mathbf{H}[g](x) \equiv \mathcal{H}[g]\Bigl(\theta[g](x),\vec{x}\Bigr) \; .
\label{fullH}
\end{equation}

We define the graviton field $h_{\mu\nu}(x)$ as the perturbation of the 
conformally rescaled metric about de Sitter background,
\begin{equation}
g_{\mu\nu}(x) \equiv a^2(\eta) \Bigl[ \eta_{\mu\nu} + \kappa h_{\mu\nu}(x)
\Bigr] \equiv a^2(\eta) \widetilde{g}_{\mu\nu}(x) \qquad , \qquad \kappa^2 
\equiv 16 \pi G \; . \label{hdef}
\end{equation}
We adhere to the usual convention that graviton indices are raised and 
lowered with the Lorentz metric, for example, $h^{\mu\nu} \equiv \eta^{\mu\rho}
\eta^{\nu\sigma} h_{\rho\sigma}$. The Feynman rules are given in terms of 
the graviton field \cite{Tsamis:1992xa,Woodard:2004ut}, so we must expand
$\mathbf{H}[g](x)$ in powers of it in order to evaluate its expectation value. 
It is useful to also expand the scalar $\Phi[g](x)$ \cite{Tsamis:2013cka},
\begin{eqnarray}
\Phi[g](x) & = & \Phi_0(\eta) + \kappa \Phi_1(x) + \kappa^2 \Phi_2(x) + 
\dots \; , \\
\Phi_1 & = & \frac1{D_A} \Biggl[-(D \!-\! 1) H h_{00} - \frac{h_{00}'}{2a}
+ \frac{h_{0j , j}}{a} - \frac{h_{jj}'}{2a} \Biggr] \; , \label{Phi1}
\end{eqnarray}
where Latin letters from the middle of the alphabet denote spatial indices,
$h \equiv \eta^{\mu\nu} h_{\mu\nu}$, a prime stands for differentiation with
respect to $\eta$ and $D_A \equiv \frac1{a^D} \partial_{\mu} (a^{D-2} 
\eta^{\mu\nu} \partial_{\nu})$. The expansion observable has a similar expansion 
\cite{Tsamis:2013cka},
\begin{eqnarray}
\mathbf{H}[g](x) & = & H + \kappa \mathbf{H}_1(x) + \kappa^2 \mathbf{H}_2(x)
+ \dots \; , \label{Hexp} \\
\mathbf{H}_1 & = & \frac12 H h_{00} + \frac{h'_{ii}}{2 (D \!-\! 1) a} +
\partial_i \Biggl[ \frac{- h_{0i}}{(D \!-\! 1) a} + \frac{\partial_i 
\Phi_1}{(D \!-\! 1) a^2} \Biggr] \; . \label{H1} \qquad 
\end{eqnarray}
The homogeneity and isotropy of our state and our gauge (see section \ref{feynman})
mean that we must get zero for the expectation value of a total spatial derivative 
such as the final term of (\ref{H1}). We will therefore not bother about giving 
such terms for $\mathbf{H}_2$ although they have been worked out 
\cite{Tsamis:2013cka},
\begin{eqnarray}
\lefteqn{\mathbf{H}_2 = \frac38 H h_{00} h_{00} - \frac12 H h_{0i} h_{0i} 
+ \frac{[ -h_{ij} h_{ij}' + h_{00 , i} h_{0i} + \frac12 h_{00} h_{ii}' - 
h_{jj , i} h_{0i} ]}{2 (D \!-\! 1) a} } \nonumber \\
& & \hspace{0cm} + \frac{[(D \!-\! 1) H a h_{00}' + (\partial_0^2 \!-\! H a 
\partial_0 \!-\! \nabla^2) h_{ii} - 2 (\partial_0 \!-\! H a) h_{0i,i} + h_{00} 
\nabla^2 ] \Phi_1}{2 (D \!-\! 1) a^2} \nonumber \\
& & \hspace{2cm} + \Bigl( \frac{D \!+\! 1}{D \!-\! 1}\Bigr) \frac{ H \partial_i
\Phi_1 \partial_i \Phi_1}{2 a^2} + \Bigl({\rm Spatial\ Derivative\ Terms}
\Bigr) \; . \qquad \label{H2}
\end{eqnarray}

Our expansion observable $\mathbf{H}[g](x)$ is a nonlocal composite operator 
functional of the metric, which makes its renormalization problematic. The 
normal BPHZ (Bogoliubov and Parasiuk \cite{Bogoliubov:1957gp}, Hepp
\cite{Hepp:1966eg} and Zimmerman \cite{Zimmermann:1968mu,Zimmermann:1969jj})
renormalization technique only suffices to remove ultraviolet divergences 
from noncoincident 1PI (one-particle-irreducible) functions. It is known how
to perform additional renormalizations to remove the divergences of local
composite operators \cite{Itzykson:1980rh,Weinberg:1996kr}. However, the
only nonlocal composite operator whose renormalization we now understand
is the Wilson loop of non-Abelian gauge theory \cite{Korchemsky:1987wg}.
 
A way forward may be the observation that our expansion observable can be
considered as descending from a local composite operator in the scalar-metric
theory whose Lagrangian is,
\begin{equation}
\mathcal{L} = \frac1{16 \pi G} \Bigl[ R - (D\!-\! 2) \Lambda\Bigr]
\sqrt{-g} - \frac12 \partial_{\mu} \varphi \partial_{\nu} \varphi g^{\mu\nu} 
\sqrt{-g} - (D \!-\! 1) H \varphi \sqrt{-g} \; . \label{matterL}
\end{equation}
Here the full cosmological constant is $\Lambda \equiv (D-1) H^2 + 
\delta \Lambda$. Although the scalar $\varphi$ obeys the same equation 
(\ref{Phieqn}) as $\Phi[g]$, it possesses its own independent initial value
data instead of being completely fixed by the initial conditions (\ref{IC}).
Nonetheless, we conjecture that the composite operator renormalization of 
$\mathbf{H}[g](x)$ may be the same as the composite operator renormalization
of the corresponding operator in (\ref{matterL}). We will see in 
section~\ref{renorm} that this conjecture is correct, at least at one loop
order.

\section{One Loop Expectation Value} \label{1loop}

The purpose of this section is to evaluate the expectation value of 
$\mathbf{H}[g](x)$ at one loop order. We begin with the Feynman rules.
Next the 1-point contribution is inferred from previous work and the various 
2-point contributions are reduced to convolutions of propagators. These 
convolutions are then reduced to the coincidence limits of integrated 
propagators whose evaluation is explained in Appendices A-C.

\subsection{Feynman Rules} \label{feynman}

The invariant part of the pure gravitational action can be expressed in terms
of the fields $h_{\mu\nu}$ and $\widetilde{g}_{\mu\nu} = \eta_{\mu\nu} + 
\kappa h_{\mu\nu}$ defined in expression (\ref{hdef}) as \cite{Tsamis:1992xa},
\begin{eqnarray}
\lefteqn{\mathcal{L} = \frac1{16 \pi G} \, \Bigl[R - (D\!-\!2) \Lambda\Bigr]
\sqrt{-g} + {\rm Counterterms} \; , } \\
& & = \Bigl({\rm Surface\ Terms}\Bigr) + \frac12 (D \!-\! 2) H a^{D-1} 
\sqrt{-\widetilde{g}} \, \widetilde{g}^{\rho\sigma} \widetilde{g}^{\mu\nu} 
h_{\rho\sigma ,\mu} h_{\nu 0} \nonumber \\
& & \hspace{1.5cm} + a^{D-2} \sqrt{-\widetilde{g}} \, \widetilde{g}^{\alpha\beta} 
\widetilde{g}^{\rho\sigma} \widetilde{g}^{\mu\nu} \Biggl\{ \frac12 
h_{\alpha\rho ,\mu} h_{\nu\sigma ,\beta} \!-\! \frac12 h_{\alpha\beta ,\rho} 
h_{\sigma\mu ,\nu} \nonumber \\
& & \hspace{3.9cm} + \frac14 h_{\alpha\beta ,\rho} h_{\mu\nu ,\sigma} 
\!-\! \frac14 h_{\alpha\rho ,\mu} h_{\beta\sigma ,\nu} \Biggr\} + 
{\rm Counterterms} \; . \qquad \label{invL}
\end{eqnarray}
Because the expansion observable $\mathbf{H}[g](x)$ is gauge invariant it 
should not matter how we fix the gauge, so we make the choice which gives the
simplest propagator. That choice is defined by adding the gauge fixing term
\cite{Tsamis:1992xa,Woodard:2004ut},
\begin{equation}
\mathcal{L}_{\rm GF} = -\frac12 a^{D-2} \eta^{\mu\nu} F_{\mu} F_{\nu} 
\quad , \quad
F_{\mu} \equiv \eta^{\rho\sigma} \Bigl[ h_{\mu \rho , \sigma} - \frac12
h_{\rho \sigma , \mu} + (D - 2) a H h^{\mu \rho} \delta^0_{\sigma}
\Bigr] \; . \label{Lgauge}
\end{equation}
The associated ghost Lagrangian (with anti-ghost field $\gamma^{\mu}(x)$ and 
ghost field $\epsilon_{\sigma}(x)$) is \cite{Tsamis:1992xa},
\begin{eqnarray}
\lefteqn{\mathcal{L}_{gh} = -2 a^{D-2} \gamma^{\mu}_{~ ,\alpha} 
\eta^{\alpha\beta} \eta^{\rho\sigma} \Bigl[ \widetilde{g}_{\rho ( \mu} 
\partial_{\beta )} + \frac12 \widetilde{g}_{\mu \beta , \rho} + H a 
\widetilde{g}_{\mu\beta} \delta^0_{~\rho} \Bigr] \epsilon_{\sigma} }
\nonumber \\
& & \hspace{2.5cm} + (a^{D-2} \gamma^{\mu})_{,\mu} \eta^{\alpha\beta} 
\eta^{\rho\sigma} \Bigl[ \widetilde{g}_{\rho \alpha} \partial_{\beta} + 
\frac12 \widetilde{g}_{\alpha\beta , \rho} + H a \widetilde{g}_{\alpha\beta} 
\delta^0_{~\rho} \Bigr] \epsilon_{\sigma} \; , \label{Lghost} \qquad 
\end{eqnarray}
where parenthesized indices are symmetrized.

In the gauge (\ref{Lgauge}) both the graviton and ghost propagators are 
simple because they consist of sums of known scalar propagators, each 
multiplied by an index factor which is constant in space and time \cite{Tsamis:1992xa,Woodard:2004ut},
\begin{eqnarray}
i\Bigl[\mbox{}_{\mu\nu} \Delta_{\rho\sigma}\Bigr](x;x') & = & 
\sum_{I=A,B,C} i\Delta_I(x;x') \times \Bigl[\mbox{}_{\mu\nu} 
T^I_{\rho\sigma}\Bigr] \; , \label{gravprop} \qquad \\
i\Bigl[\mbox{}_{\mu} \Delta_{\nu}\Bigr](x;x') & = & i \Delta_A(x;x') 
\!\times\! \overline{\eta}_{\mu\nu} - i\Delta_B(x;x') \!\times\!
\delta^0_{~\mu} \delta^0_{~\nu} \; . \label{ghostprop} \qquad
\end{eqnarray}
Here $\eta_{\mu\nu}$ is the Lorentz metric and $\overline{\eta}_{\mu\nu}
\equiv \eta_{\mu\nu} + \delta^0_{~\mu} \delta^0_{~\nu}$ is its purely
spatial part. The same constant tensors suffice to give the three index 
factors of the graviton propagator,
\begin{eqnarray}
\Bigl[\mbox{}_{\mu\nu} T^A_{\rho\sigma}\Bigr] & = & 2 \overline{\eta}_{\mu (\rho}
\overline{\eta}_{\sigma ) \nu} - \frac{2}{D \!-\! 3} \, \overline{\eta}_{\mu\nu}
\overline{\eta}_{\rho\sigma} \; , \\
\Bigl[\mbox{}_{\mu\nu} T^B_{\rho\sigma}\Bigr] & = & - 4 \delta^0_{( \mu} 
\overline{\eta}_{\nu ) (\rho} \delta^0_{\sigma )} \; , \\
\Bigl[\mbox{}_{\mu\nu} T^C_{\rho\sigma}\Bigr] & = & \frac{2}{(D \!-\! 2) (D \!-\! 3)}
\Bigl[ (D \!-\! 3) \delta^0_{\mu} \delta^0_{\nu} \!+\! \overline{\eta}_{\mu\nu}
\Bigr] \Bigl[ (D \!-\! 3) \delta^0_{\rho} \delta^0_{\sigma} \!+\! 
\overline{\eta}_{\rho\sigma} \Bigr] \; .
\end{eqnarray}

The three scalar propagators have masses $M_A^2 = 0$, $M_B^2 = (D-2) H^2$ and
$M_C^2 = 2 (D-3) H^2$. They are most easily represented in terms of the de
Sitter length function $y(x;x')$,
\begin{equation}
y(x;x') \equiv H^2 a(\eta) a(\eta') \Biggl[ \Bigl\Vert \vec{x} \!-\! \vec{x}' 
\Bigr\Vert^2 - \Bigl( \vert \eta \!-\! \eta'\vert - i \epsilon\Bigr)^2 \Biggr] 
\; . \label{ydef}
\end{equation}
The $A$-type scalar is well known to break de Sitter invariance 
\cite{Onemli:2002hr,Onemli:2004mb},
\begin{eqnarray}
\lefteqn{i \Delta_A = \frac{H^{D-2}}{(4\pi)^{\frac{D}2}} \Biggl\{
\frac{\Gamma(\frac{D}2)}{\frac{D}2 \!-\! 1} \Bigl( \frac{4}{y}\Bigr)^{\!\frac{D}2-1}
\!\!\!\!\!\!\!+\! \frac{\Gamma(\frac{D}2 \!+\! 1)}{\frac{D}2 \!-\! 2} 
\Bigl(\frac{4}{y}\Bigr)^{\!\frac{D}2-2} \!\!\!\!\!\!\!-\! 
\frac{\Gamma(D\!-\!1)}{\Gamma(\frac{D}2)} \Biggl[ \pi \cot\Bigl(\pi 
\frac{D}{2}\Bigr) \!-\! \ln(a a')\Biggr] } \nonumber \\
& & \hspace{1cm} + \sum_{n=1}^{\infty} \Biggl[
\frac1{n} \frac{\Gamma(n \!+\! D \!-\! 1)}{\Gamma(n \!+\! \frac{D}2)}
\Bigl(\frac{y}4 \Bigr)^n \!\!\!\! - \frac1{n \!-\! \frac{D}2 \!+\! 2}
\frac{\Gamma(n \!+\!  \frac{D}2 \!+\! 1)}{\Gamma(n \!+\! 2)} \Bigl(\frac{y}4
\Bigr)^{n - \frac{D}2 +2} \Biggr] \Biggr\} . \qquad \label{DeltaA}
\end{eqnarray}
On the other hand, the $B$-type and $C$-type propagators are de Sitter invariant
functions of $y(x;x')$,
\begin{equation}
i \Delta_B = \frac{H^{D-2}}{(4\pi)^{\frac{D}2}} \Biggl\{
\frac{\Gamma(\frac{D}2)}{\frac{D}2 \!-\! 1} \Bigl( \frac{4}{y}\Bigr)^{\frac{D}2-1}
\!\!\!\!\!+\! \sum_{n=0}^{\infty} \Biggl[ \frac{\Gamma(n \!+\! 
\frac{D}2)}{\Gamma(n \!+\!2)} \Bigl( \frac{y}4 \Bigr)^{n - \frac{D}2 +2} 
\!\!\!\!\!-\! \frac{\Gamma(n \!+\! D \!-\! 2)}{\Gamma(n \!+\! \frac{D}2)} 
\Bigl(\frac{y}4 \Bigr)^n \Biggr] \Biggr\} , \label{DeltaB}
\end{equation}
\begin{eqnarray}
\lefteqn{i \Delta_C = \frac{H^{D-2}}{(4\pi)^{\frac{D}2}} \Biggl\{ 
\frac{\Gamma(\frac{D}2)}{\frac{D}2 \!-\! 1} \Bigl( \frac{4}{y}\Bigr)^{\frac{D}2-1}
\!\!\!\!- \sum_{n=0}^{\infty} \Biggl[ \Bigl(n \!-\! \frac{D}2 \!+\!  3\Bigr) 
\frac{\Gamma(n \!+\! \frac{D}2 \!-\! 1)}{\Gamma(n \!+\! 2)} \Bigl(\frac{y}4
\Bigr)^{n - \frac{D}2 +2} } \nonumber \\
& & \hspace{6.5cm} - (n\!+\!1) \frac{\Gamma(n \!+\! D \!-\! 3)}{\Gamma(n \!+\! 
\frac{D}2)} \Bigl(\frac{y}4 \Bigr)^n \Biggr] \Biggr\} . \qquad \label{DeltaC}
\end{eqnarray}
Although the infinite summations may appear daunting, each of the scalar 
propagators takes a simple form for $D = 4$ dimensions,
\begin{equation}
D =4 \quad \Longrightarrow \quad i\Delta_A = \frac{H^2}{4 \pi^2} \Biggl[ 
\frac1{y} - \frac12 \ln\Bigl( \frac{y}{4 a a'}\Bigr) \Biggr] \;\; , \;\;
i\Delta_B = i\Delta_C = \frac{H^2}{4 \pi^2 y} \; .
\end{equation}
This means we only need the sums when there is a divergence, and then only a 
few of the lowest values of $n$ are required.

We close by giving a unified treatment of differential operators and 
propagators on de Sitter background. The inverse of the massless scalar 
d'Alem\-bert\-ian $D_A$ has already appeared in expression (\ref{Phi1}) 
for $\Phi_1(x)$,
\begin{equation}
D_A \equiv \frac1{a^D} \, \partial_{\mu} \Bigl[ a^{D-2} \, \eta^{\mu\nu} 
\partial_{\nu} \Bigr] = \frac1{a^2} \, \Bigl[ \partial^2 - (D\!-\!2) H a 
\partial_0\Bigr] \; . \label{DA}
\end{equation}
We denote a general massive scalar kinetic operator with a subscript $\nu$,
\begin{equation}
D_{\nu} \equiv D_A + (\nu^2 \!-\! \nu_A^2) H^2 \qquad , \qquad 
\nu_A \equiv \Bigl( \frac{D \!-\! 1}{2} \Bigr) \; . \label{Dnu}
\end{equation}
The various propagators are obtained by acting the inverse differential 
operators on a delta function,
\begin{equation}
i\Delta_{\nu}(x;x') = \frac1{D_{\nu}} \Bigl[ i\delta^D(x \!-\! x') \Bigr]
= \frac1{D'_{\nu}} \Bigl[i\delta^D(x \!-\! x') \Bigr] \; . \label{genprop}
\end{equation}
In addition to propagators we also require some integrated propagators whose
evaluation is explained in Appendix C,
\begin{eqnarray}
I_{\alpha\beta}(x;x') \equiv \frac1{D_{\alpha}} \Bigl[ i\Delta_{\beta}(x;x')
\Bigr] & , & I_{\alpha\beta\gamma}(x;x') \equiv \frac1{D_{\alpha}} 
\frac1{D'_{\gamma}} \Bigl[i\Delta_{\beta}(x;x') \Bigr]  \; , \label{Iints} \\
J_{\alpha\beta}(x;x') \equiv \frac1{D_{\alpha}} \Bigl[
\frac{i\Delta_{\beta}(x;x')}{a(\eta)} \Bigr] & , & J_{\alpha\beta\gamma}(x;x') 
\equiv \frac1{D_{\alpha}} \frac1{D'_{\gamma}} \Bigl[ 
\frac{i\Delta_{\beta}(x;x')}{a(\eta) a(\eta')} \Bigr] \; , \label{Jints} \\
& & K_{\alpha\beta\gamma}(x;x') \equiv \frac1{D_{\alpha}} \frac1{D'_{\gamma}} 
\Bigl[\frac{i\Delta_{\beta}(x;x')}{a(\eta')} \Bigr] \; . \label{JKints} \qquad 
\end{eqnarray}
For the special case where the index corresponds to a particularly useful
propagator we have found it convenient to employ an alternate, alphabetical, 
representation according to the scheme,
\begin{equation}
\nu_B \equiv \nu_A - 1 \quad , \quad \nu_C \equiv \nu_A - 2 \quad , \quad
\nu_D \equiv \nu_A - 3 \; . \label{nuBCD}
\end{equation}
The last case, $\nu_D$, corresponds to a scalar of mass $M^2_D = 3 (D - 4) 
H^2$ which does not appear in the graviton or ghost propagators. It nonetheless
occurs in our reductions of the 2-point contributions (see section \ref{2point})
when using identities of Appendix B to reflect a time derivative from one side 
of a $C$-type propagator to the other,
\begin{equation}
\Bigl[ \partial_0' + 2 H a'\Bigr] i\Delta_C(x;x') = -\Bigl[ \partial_0 +
(D \!-\! 4) H a\Bigr] i\Delta_D(x;x') \; . \label{CtoD}
\end{equation}

\subsection{1-Point Contributions} \label{1point}

\begin{figure}[ht]
\vskip -1cm
\begin{center}
\includegraphics[width=10cm,height=3.125cm]{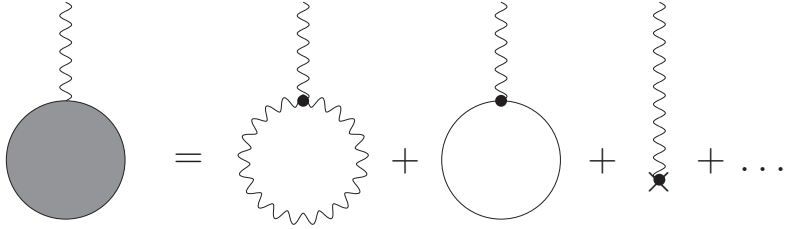}
\end{center}
\vskip 1cm
\caption{Diagrammatic representation of $\langle \Omega \vert \kappa 
\mathbf{H}_1(x) \vert \Omega \rangle$. Graviton lines are wavy and 
ghost lines are straight.}
\label{Diagram1}
\end{figure}

Figure \ref{Diagram1} shows the diagrams which contribute to the
one loop expectation value of $\kappa \mathbf{H}_1(x)$. The first two
diagrams display the primitive contribution while the third diagram gives
the contribution from the cosmological counterterm. The primitive diagrams
have been evaluated previously using dimensional regularization and in the
same gauge (\ref{Lgauge}) that we employ. The result implies,\footnote{The
inverses of $D_A$ and $D_C$ act on constants with the homogeneous solutions
($a^0$ and $a^{1-D}$ for $D_A$ and $a^{-2}$ and $a^{3-D}$ for $D_C$)
chosen so that expression (\ref{H1exp}) is constant,
\begin{eqnarray}
\frac1{D_A} \, C = -\frac{C \ln(a)}{(D \!-\! 1) H^2} \qquad , \qquad
\frac1{D_C} \, C = -\frac{C}{2 (D \!-\! 3) H^2} \; . \nonumber
\end{eqnarray}}
\begin{eqnarray}
\lefteqn{\Bigl\langle \Omega \Bigl\vert \kappa \mathbf{H}_1(x) \Bigr\vert 
\Omega \Bigr\rangle = \frac{2 H}{D \!-\! 2} \frac1{D_C} \Bigl[-K + \frac12 
(D \!-\! 2) \delta \Lambda\Bigr] } \nonumber \\
& & \hspace{2.5cm} + \frac{2 \partial_0}{(D \!-\! 2) (D\!-\!3)a} \Bigl[
\frac1{D_C} - \frac{D \!-\! 2}{D_A}\Bigr] \Bigl[-K + \frac12 (D \!-\! 2) 
\delta \Lambda\Bigr] \; , \label{H1exp} \qquad
\end{eqnarray}
where the constant $K$ is \cite{Tsamis:2005je},
\begin{equation}
K = \frac{\kappa^2 H^D}{(4 \pi)^{\frac{D}2}} \frac{\Gamma(D \!-\! 1)}{
\Gamma(\frac{D}2)} \Biggl[ \frac1{D \!-\! 3} - \frac12 (D \!-\! 2) (D \!+\! 1)
+ \frac18 (D \!-\! 4) (D \!-\! 1) \Biggr] \; . 
\end{equation}
The correct renormalization condition for $\delta \Lambda$ seems to be to 
make the trace of the graviton 1PI 1-point function vanish on the initial 
value surface. Otherwise, the constant we call ``$H$'' does not represent
the initial expansion rate. If this condition is adopted then expression 
(\ref{H1exp}) vanishes,
\begin{equation}
\delta \Lambda = \frac{2 K}{D \!-\! 2} \qquad \Longrightarrow \qquad
\Bigl\langle \Omega \Bigl\vert \kappa \mathbf{H}_1(x) \Bigr\vert \Omega 
\Bigr\rangle = 0 \; . \label{deltaLcondition}
\end{equation}

Condition (\ref{deltaLcondition}) provides for the simplest development of 
perturbation theory, however, it is worth examining what would happen if a
different renormalization condition were adopted. In this case the 
expectation value of the graviton field would not vanish at one loop order, 
and its spatial components would suffer secular growth in our gauge. Both 
the failure to vanish and the secular growth follow from having declined to 
make the parameter $H$ in the background metric agree with the true (initial)
expansion rate which we might write as $H + \delta H$. To see this, suppose 
we change (\ref{deltaLcondition}) to,
\begin{equation}
-K + \frac12 (D \!-\! 2) \delta \Lambda = (D\!-\!2) (D \!-\! 1) H \delta H
\; . \label{altLcondition}
\end{equation}
Then the two nonzero terms of expression (\ref{H1exp}) become,
\begin{eqnarray}
\frac{2 H}{D \!-\! 2} \frac1{D_C} \Bigl[ (D \!-\! 2) (D\!-\!1) H \delta H
\Bigr] & = & -\Bigl(\frac{D \!-\! 1}{D \!-\! 3}\Bigr) \delta H 
\label{1stterm} \; , \\
-\frac{2 a^{-1} \partial_0}{D \!-\! 3} \frac1{D_A} \Bigl[ (D \!-\!2)
(D \!-\! 1) H \delta H \Bigr] & = & 2 \Bigl( \frac{D \!-\! 2}{D \!-\! 3}
\Bigr) \delta H \; . \label{2ndterm}
\end{eqnarray}
The sum of (\ref{1stterm}) and (\ref{2ndterm}) gives precisely $\delta H$,
which makes for a nice check on the consistency of our expansion observable.
Of course most researchers would at this stage absorb $\delta H$ into $H$
so as to make condition (\ref{deltaLcondition}) pertain. Persisting with a 
nonzero value of $\delta H$ would be like working in flat space QED 
(quantum electrodynamics) with the parameter $m$ failing to stand for the 
actual electron mass.

\subsection{2-Point Contributions} \label{2point}

\begin{figure}[ht]
\vskip -1cm
\begin{center}
\includegraphics[width=6.25cm,height=3.125cm]{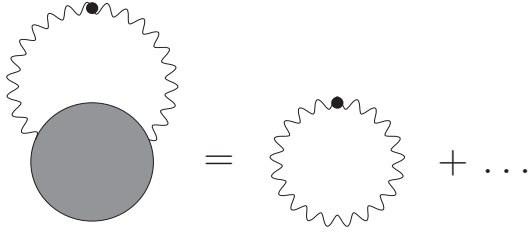}
\end{center}
\vskip 1cm
\caption{Diagrammatic representation of $\langle \Omega \vert \kappa^2 
\mathbf{H}_2(x) \vert \Omega \rangle$. Graviton lines are wavy and 
ghost lines are straight.}
\label{Diagram2}
\end{figure} 

The simple diagrammatic structure of the one loop 2-point contributions
which is shown in Figure \ref{Diagram2} conceals the enormous complexity
of our observable. One can see from expression (\ref{H2}) that $\mathbf{H}_2$ 
contains three distinct classes of terms: those with both graviton fields at 
the point $x^{\mu}$, those with one graviton at $x^{\mu}$ and the other acted 
upon by a factor of $\frac1{D_A}$, and those with both gravitons acted upon 
by (different) factors of $\frac1{D_A}$. In each case we substitute expression 
(\ref{gravprop}) and then perform the indicated tensor contractions and 
differentiations. However, factors of $\frac1{D_A}$ require special treatment 
to reflect all the derivatives outside the inverse differential operators. We 
will explicitly work out a sample reduction from each of the three classes, 
and then give the full result for that class.

To illustrate the reduction of the first class we have selected the third term 
on the first line of expression (\ref{H2}). Multiplying by the factor of 
$\kappa^2$ which all parts of $\mathbf{H}_2$ carry gives,
\begin{eqnarray}
\lefteqn{ \Bigl\langle \Omega \Bigl\vert -\frac{\kappa^2 h_{ij}(x) h'_{\ij}(x)}{
2 (D \!-\! 1) a} \Bigr\vert \Omega \Bigr\rangle = -\frac{\kappa^2}{2 (D \!-\! 1) a}
\lim_{x' \rightarrow x} \Biggl\{ \Bigl[ \delta_{ii} \delta_{jj} + \delta_{ij}
\delta_{ij} } \nonumber \\
& & \hspace{.5cm} - \frac2{D \!-\! 3} \, \delta_{ij} \delta_{ij} \Bigr] \partial_0'
i\Delta_A(x;x') + \frac{2}{(D \!-\! 3) (D \!-\! 2)} \, \delta_{ij} \delta_{ij}
\partial_0' i\Delta_C(x;x') \Biggr\} , \qquad \\
& & \hspace{-.5cm} = -\frac{\kappa^2}{2 a} \lim_{x' \rightarrow x} \Biggl\{
\Bigl[D - \frac{2}{D \!-\! 3}\Bigr] \partial_0' i\Delta_A(x;x') + \frac{2}{
(D \!-\! 3) (D \!-\! 2)} \, \partial'_0 i\Delta_C(x;x') \Biggr\} , \qquad \\
& & \hspace{-.5cm} = -\frac{\kappa^2}{2} \Bigl[D - \frac{2}{D \!-\! 3}\Bigr] 
\frac{H^{D-2}}{(4\pi)^{\frac{D}2}} \, \frac{\Gamma(D \!-\! 1)}{\Gamma(\frac{D}2)}
\, H  = -\frac{\kappa^2 H^3}{8 \pi^2} + O(D \!-\! 4) \; . \qquad 
\end{eqnarray} 
The full result for this class of terms is,
\begin{eqnarray}
\lefteqn{\Bigl\langle \Omega \Bigl\vert \Bigl[ \kappa^2 \mathbf{H}_2(x) 
\Bigr]_{hh} \Bigr\vert \Omega \Bigr\rangle = 
\frac{\kappa^2 H^{D-1}}{(4\pi)^{\frac{D}2}} 
\frac{\Gamma(D\!-\!2)}{\Gamma(\frac{D}2)} \Bigg\{ \frac34 \frac1{D \!-\! 2} 
-\frac12 (D \!-\!1) } \nonumber \\
& & \hspace{2cm} - \frac12 (D\!-\!2) \Bigl(D - \frac{2}{D\!-\!3}\Bigr) \Biggr\} 
\longrightarrow -\frac{25}{8} \!\times\! \frac{\kappa^2 H^3}{16 \pi^2} + 
O(D \!-\! 4) \; . \qquad \label{hhterms}
\end{eqnarray}

The second class consists of one local $h_{\mu\nu}(x)$ and the other inside 
a first order correction of the scalar (\ref{Phi1}). For example, consider the 
second term on the second line of expression (\ref{H2}) with the last of the 
four terms of $\Phi_1(x)$,
\begin{equation}
\frac{\kappa^2}{2 (D \!-\! 1) a^2} \!\times\! \Bigl( \partial_0^2 \!-\! H a 
\partial_0 \!-\! \nabla^2\Bigr) h_{ii}(x) \!\times\! \frac1{D_A} \Bigl[ -
\frac{h_{jj}'(x)}{2 a} \Bigr] \; . \label{2ndexamp}
\end{equation}
We write the expectation value of any such term as the coincidence limit of 
$\frac1{D_A}$ acting on a differentiated propagator. Then we employ the 
reflection identities of Appendix B to move derivatives outside the factor of 
$\frac1{D_A}$, and the result is expressed in terms of the integrated scalar 
propagators (\ref{Iints}-\ref{Jints}). For expression (\ref{2ndexamp}) the 
reduction is,
\begin{eqnarray}
\lefteqn{\Bigl\langle \Omega \Bigl\vert {\rm Exp}(\ref{2ndexamp}) 
\Bigr\vert \Omega \Bigr\rangle = -\kappa^2 \lim_{x' \rightarrow x} 
\frac{( \partial^2_0 \!-\! H a \partial_0 \!-\! \nabla^2)}{4 (D \!-\! 1) a^2} 
\frac1{D_A'} \Biggl\{ \frac1{a'} \partial_0' i\Bigl[\mbox{}_{ii} 
\Delta_{jj}\Bigr](x;x') \Biggr\} , } \\
& & \hspace{-.5cm} = \kappa^2 \lim_{x' \rightarrow x} \frac{( \partial^2_0 \!-\! 
H a \partial_0 \!-\! \nabla^2)}{(D \!-\! 3) a^2} \frac1{D_A'} \Biggl\{  
\frac{\partial_0' i\Delta_A(x;x')}{a'} - \Bigl( \frac{D \!-\! 1}{D \!-\! 2}\Bigr) 
\frac{\partial_0' i\Delta_C(x;x')}{2 a'} \Biggr\} , \qquad \\
& & \hspace{-.5cm} = \kappa^2 \lim_{x' \rightarrow x} \frac{( \partial^2_0 \!-\! 
H a \partial_0 \!-\! \nabla^2)}{(D \!-\! 3) a^2} \frac1{D_A'} \Biggl\{  -\Bigl[
\partial_0 + (D \!-\! 2) H a\Bigr] \frac{i\Delta_B(x;x')}{a'} \nonumber \\
& & \hspace{.5cm} + \Bigl( \frac{D \!-\! 1}{D \!-\! 2}\Bigr) H i\Delta_C(x;x')
+ \frac12 \Bigl( \frac{D \!-\! 1}{D \!-\! 2}\Bigr) \Bigl[ \partial_0 + 
(D \!-\! 4) H a\Bigr] \frac{ i\Delta_D(x;x')}{a'} \Biggr\} , \qquad \\
& & \hspace{-.5cm} = \kappa^2 \lim_{x' \rightarrow x} \frac{( \partial^2_0 \!-\! 
H a \partial_0 \!-\! \nabla^2)}{(D \!-\! 3) a^2} \Biggl\{  -\Bigl[\partial_0 + 
(D \!-\! 2) H a\Bigr] J_{AB}(x';x) \nonumber \\
& & \hspace{.5cm} + \Bigl( \frac{D \!-\! 1}{D \!-\! 2}\Bigr) H I_{AC}(x';x) 
+ \frac12 \Bigl( \frac{D \!-\! 1}{D \!-\! 2}\Bigr) \Bigl[ \partial_0 + 
(D \!-\! 4) H a\Bigr] J_{AD}(x';x) \Biggr\} . \qquad 
\end{eqnarray}
Using Appendix C the full result for this class of terms is,
\begin{eqnarray}
\lefteqn{\Bigl\langle \Omega \Bigl\vert \Bigl[ \kappa^2 \mathbf{H}_2(x) 
\Bigr]_{h\Phi_1} \Bigr\vert \Omega \Bigr\rangle = \frac{\kappa^2}{a^2}
\lim_{x' \rightarrow x} \Biggl\{ \Bigl[\partial_0^2 + (D\!-\!4) H a \partial_0
- \frac{2 \nabla^2}{D \!-\! 1} \Bigr] } \nonumber \\
& & \hspace{0cm} \times \Biggl[ -\frac{(D^2 \!-\! 6 D \!+\! 7)}{(D \!-\! 2)
(D \!-\! 3)} \, H I_{AC}(x';x) + \Bigl[\partial_0 + (D \!-\! 4) H a\Bigr]
\frac{J_{AD}(x';x)}{D \!-\! 3} \Biggr] \nonumber \\
& & \hspace{.5cm} - \Biggl[ \frac{(\partial_0^2 \!-\! H a \partial_0 \!-\! 
\nabla^2) [ \partial_0 \!+\! (D \!-\! 2) H a]}{D \!-\! 3} + \frac{ \nabla^2 
( \partial_0 \!-\! H a)}{D \!-\! 1} \Biggr] J_{AB}(x';x) \Biggr\} \qquad
\nonumber \\
& & \longrightarrow \frac{\kappa^2 H^{D-1}}{(4\pi)^{\frac{D}2}} 
\frac{\Gamma( D \!-\! 1)}{\Gamma(\frac{D}2)} \Biggl\{ 
\frac{-\frac{132193}{5005}}{D \!-\! 4} + O\Bigl( (D \!-\!4)^0\Bigr) \Biggr\} . 
\qquad \label{hPhiterms}
\end{eqnarray}

The final class of terms consists of a graviton from one factor of $\Phi_1(x)$
multiplied by a graviton from another factor of $\Phi_1(x)$. As an example, 
consider the case where it is the second graviton of expression (\ref{Phi1}) 
which is contributed by each $\Phi_1(x)$,
\begin{equation}
\Bigl( \frac{D \!+\! 1}{D \!-\! 1} \Bigr) \frac{\kappa^2 H}{2 a^2} 
\frac{\partial_i}{D_A} \Bigl[ -\frac{h_{00}'(x)}{2 a}\Bigr] \times
\frac{\partial_i}{D_A} \Bigl[ -\frac{h_{00}'(x)}{2 a}\Bigr] \; . \label{3rdexamp}
\end{equation}
The reduction of expression (\ref{3rdexamp}) proceeds similarly to that of
(\ref{2ndexamp}), 
\begin{eqnarray}
\lefteqn{\Bigl\langle \Omega \Bigl\vert {\rm Exp}(\ref{3rdexamp}) \Bigr\vert 
\Omega \Bigr\rangle = \frac{(D\!+\!1) (D \!-\! 3)}{(D\!-\!1) (D\!-\!2)} 
\frac{\kappa^2 H}{4 a^2} \lim_{x' \rightarrow x} \frac{ \partial_i}{D_A} 
\frac{\partial_i'}{D_A'} \Biggl\{ \frac{\partial_0 \partial_0' 
i\Delta_C(x;x')}{a a'} \Biggr\} ,} \\
& & \hspace{-.5cm} = -\frac{(D\!+\!1) (D \!-\! 3)}{(D\!-\!1) (D\!-\!2)} 
\frac{\kappa^2 H}{4 a^2} \lim_{x' \rightarrow x} \frac{ \nabla^2}{D_A D_A'} 
\Biggl\{ \Bigl( \partial_0 \!+\! H a\Bigr) \Bigl( \partial_0' \!+\! H a'\Bigr)
\frac{i \Delta_C(x;x')}{a a'} \Biggr\} , \qquad \\
& & \hspace{-.5cm} = -\frac{(D\!+\!1) (D \!-\! 3)}{(D\!-\!1) (D\!-\!2)} 
\frac{\kappa^2 H}{4 a^2} \lim_{x' \rightarrow x} \Biggl\{2 H \Bigl[ \partial_0' 
+ (D \!-\! 2) H a'\Bigr] \nabla^2 K_{ACB}(x;x') \nonumber \\
& & \hspace{-.5cm} + \Bigl[ \partial_0 \!+\! (D \!-\! 2) H a\Bigr] \Bigl[ 
\partial_0' \!+\! (D \!-\! 2) H a'\Bigr] \nabla^2 J_{BCB}(x;x') \!+\! H^2 
\nabla^2 I_{ACA}(x;x') \Biggr\} . \qquad
\end{eqnarray}
With Appendix C we find that the full result for this class of terms is,
\begin{eqnarray}
\lefteqn{\Bigl\langle \Omega \Bigl\vert \Bigl[ \kappa^2 \mathbf{H}_2(x)
\Bigr]_{\Phi_1 \Phi_1} \Bigr\vert \Omega \Bigr\rangle = \Bigl( 
\frac{D \!+\! 1}{D \!-\! 1}\Bigr) \frac{\kappa^2 H}{2 a^2} \lim_{x' 
\rightarrow x} \nabla^2 \Biggl\{ \Bigl(\frac{D \!-\! 1}{D \!-\! 3}\Bigr) 
\, H^2 I_{AAA}(x;x') } \nonumber \\
& & \hspace{0cm} - 2 \frac{(D^2 \!-\! 3D \!+\! 1)^2}{(D\!-\!3) (D\!-\!2)}
\, H^2 I_{ACA}(x;x') -\nabla^2 J_{ABA}(x;x') + \Bigl[ \partial_0 \!+\! 
(D \!-\! 2) H a\Bigr] \nonumber \\
& & \hspace{1cm} \times \Bigl[ \partial_0' \!+\! (D \!-\! 2) H a'\Bigr] 
\Biggl[ \Bigl( \frac{D \!-\!1}{D\!-\!3}\Bigr) J_{BAB}(x;x') -2 \Bigl( 
\frac{D \!-\! 2}{D \!-\! 3}\Bigr) J_{BCB}(x;x') \Biggr] \nonumber \\
& & \hspace{-.7cm} + H \Bigl[ \partial_0' \!+\! (D \!-\!2) H a'\Bigr] 
\Biggl[2 \Bigl( \frac{D\!-\!1}{D\!-\!3} \Bigr) K_{AAB}(x;x') \!-\! 
4 \Bigl(\frac{D^2 \!-\! 3D \!+\! 1}{D \!-\! 3}\Bigr) K_{ACB}(x;x') \Biggr]
\Biggr\} \nonumber \\
& & \longrightarrow \frac{\kappa^2 H^{D-1}}{(4\pi)^{\frac{D}2}} 
\frac{\Gamma( D \!-\! 1)}{\Gamma(\frac{D}2)} \Biggl\{
\frac{-\frac{2669}{288}}{D \!-\! 4} + O\Bigl( (D \!-\!4)^0\Bigr) \Biggr\} 
. \label{PhiPhiterms} 
\end{eqnarray}

Combining the results of expressions (\ref{hhterms}), (\ref{hPhiterms}) 
and (\ref{PhiPhiterms}) gives the full one loop result,
\begin{equation}
\Bigl\langle \Omega \Bigl\vert \kappa^2 \mathbf{H}_2(x) \Bigr\vert \Omega 
\Bigr\rangle = \frac{\kappa^2 H^{D-1}}{(4\pi)^{\frac{D}2}} 
\frac{\Gamma( D \!-\! 1)}{\Gamma(\frac{D}2)} \Biggl\{ 
\frac{-\frac{3 \cdot 853 \cdot 60293}{32 \cdot 13!!}}{D \!-\! 4} + 
O\Bigl( (D \!-\!4)^0\Bigr) \Biggr\} . \label{1loopH2}
\end{equation}
The divergent part of expression (\ref{1loopH2}) is what chiefly concerns
us but we take note of the fact that the finite part is also independent
of time.

\section{Renormalization} \label{renorm}

The expansion observable $\mathbf{H}[g](x)$ is a nonlocal composite
operator whose divergences are not automatically absorbed by the BPHZ 
renormalization of non-coincident 1PI functions. Indeed, we saw in 
section \ref{1point} that the natural renormalization condition for
the cosmological counterterm $\delta \Lambda$ is to cancel the initial
value of the trace of the graviton 1PI 1-point function. Because the
graviton 1PI 1-point function is a pure trace \cite{Tsamis:2005je}
this renormalization condition has the effect of completely cancelling 
the 1-point contributions of section \ref{1point}, leaving the 2-point 
contributions of section \ref{2point} unaffected. From  expression 
(\ref{1loopH2}) we see that these terms diverge at one loop order.

It should be noted that no other BPHZ renormalizations can affect the
expectation value of $\mathbf{H}[g](x)$ at one loop order. Renormalizing 
$R$ and $R^2$ is degenerate with $\Lambda$ for de Sitter background at 
this order, and the Weyl-squared term makes no contribution at all for de 
Sitter background at this order. The additional divergences of expression 
(\ref{1loopH2}) derive from the fact that $\mathbf{H}[g](x)$ is a 
composite operator, and they require composite operator renormalization.
Because $\mathbf{H}(x)$ goes like an inverse length, and the loop counting
parameter $\kappa^2$ goes like a length squared, we require operators of
dimension ${\rm length}^{-3}$ (times $\kappa^2$) which can mix with 
$\mathbf{H}[g](x)$. Had we been dealing with a {\it local} composite 
operator the list of candidates would be short, but the number of {\it 
nonlocal} candidates is infinite.

We propose that the conundrum should be resolved by limiting candidates 
to those which are local in the scalar + gravity theory (\ref{matterL})
from which $\mathbf{H}[g](x)$ descends. With this conjecture there are just 
two candidate mixing operators of the required dimension,\footnote{We have
omitted $\dot{\mathbf{H}}[g](x)$ from the list (\ref{mixingops}) because it 
vanishes on de Sitter background.}
\begin{equation}
\mathbf{O}_1[g](x) \equiv \kappa^2 R\Bigl( \theta[g](x),\vec{x}\Bigr) 
\!\times\! \mathbf{H}[g](x) \quad , \quad \mathbf{O}_2[g](x) \equiv 
\kappa^2 \mathbf{H}^3[g](x) \; , \label{mixingops}
\end{equation}
where we recall that $\theta[g](x)$ defines the surface of simultaneity
on which the scalar $\Phi[g]$ takes its background value (\ref{time}).
Both operators are proportional to $\kappa^2 H^3$ for de Sitter, so either 
can be used to completely cancel the one loop correction (\ref{1loopH2}) 
to the expectation value of $\mathbf{H}[g](x)$. That is, we think of the
renormalized expansion operator as,
\begin{equation}
\mathbf{H}[g] + \delta \mathbf{H}[g] = \mathbf{H}[g] + \mu^{D-4} \Bigl(
\alpha_1 \mathbf{O}_1[g] + \alpha_2 \mathbf{O}_2[g]\Bigr) + 
{\rm higher\ loops} \; , \label{renormH}
\end{equation}
where $\alpha_1$ and $\alpha_2$ are functions of $D$.

It remains to discuss two issues, the first of which is the finite part
of the expectation value of (\ref{renormH}) at one loop order. We {\it 
must} choose the coefficients $\alpha_1$ and $\alpha_2$ to cancel the 
divergent part, but it might be that the finite part remains nonzero and 
represents an interesting prediction of quantum gravity. This is not so for 
three reasons. First, there is no unambiguous definition of ``the finite 
part'' of the primitive expectation value (\ref{1loopH2}). For example, had
the multiplicative factors of $(4\pi)^{-\frac{D}2}$ and 
$\Gamma(D-1)/\Gamma(\frac{D}2)$ been evaluated at $D=4$, what we call the 
finite part would change. Second, the finite parts of $\alpha_1$ and 
$\alpha_2$ are equally ambiguous for the same reason. Finally, the point of 
$\mathbf{H}[g](x) + \delta \mathbf{H}[g](x)$ is to measure the spacetime 
expansion rate. If a completely arbitrary choice makes this rate fail to 
agree with $H$, even on the initial value surface, {\it and after we have 
made the graviton 1-point function vanish} (at one loop order), then we 
have failed to properly define $\mathbf{H}[g](x) + \delta \mathbf{H}[g](x)$. 
We must make its initial expectation value agree with $H$, just as we must 
make what we call ``the physical electron mass'' agree with its observed 
value. The legitimate prediction of quantum gravity is how the expectation 
value of $\mathbf{H}[g](x) + \delta \mathbf{H}[g](x)$ changes with time. 
Because the result (\ref{1loopH2}) of the primitive one loop diagrams is 
constant, as are the mixing operators (\ref{mixingops}), there is no change 
at one loop order. We do not expect that to remain true at two loop order, 
but this is contingent on the primitive two loop contributions showing 
secular growth.

The second issue is how to renormalize $\mathbf{H}[g](x)$ on more general
backgrounds than de Sitter. One must first understand that $\mathbf{H}[g](x)$
was defined to apply for a homogeneous and isotropic background. One can
see this from the fact that the spatial position has not been invariantly
fixed \cite{Tsamis:2013cka}. Had the initial state possessed spatial 
structure this could have --- and would have --- been used to modify 
$\mathbf{H}[g](x)$ so as to invariantly fix the spatial position. 

For pure gravity with a positive cosmological constant, de Sitter is the
unique homogeneous and isotropic solution. However, it is simple to add a
scalar whose background evolution supports a more general (FRW) homogeneous 
and isotropic background. The propagators and vertices for this theory are
known \cite{Iliopoulos:1998wq,Abramo:1998hj} and the computation we have
just completed could be repeated for a general $H(t) \equiv a'/a^2$. 
Because ultraviolet divergences are local, we can be confident that the
result would be divergences proportional to two terms: $H(t) \dot{H}(t)$
and $H^3(t)$. As it happens, the two one loop mixing operators 
(\ref{mixingops}) span this 2-dimensional space of possible divergences,
\begin{equation}
\mathbf{O}_1[{\rm FRW}](x) = \kappa^2 \Bigl[2 (D\!-\! 1) H \dot{H} +
D (D \!-\! 1) H^3\Bigr] \;\; , \;\; \mathbf{O}_2[{\rm FRW}](x) = \kappa^2
H^3 \; . \label{FRWops}
\end{equation}
It therefore seems inevitable that we can not only renomormalize the one
loop expectation value of $\mathbf{H}[g](x)$ on de Sitter background but
also on an arbitrary homogeneous and isotropic background.

\section{Epilogue} \label{epi}

Our task has been to give an invariant quantification of the prediction
that there is no one loop back-reaction in pure quantum gravity on de
Sitter background \cite{Tsamis:1996qq,Tsamis:2011ep}. In section \ref{expop} 
we reviewed the nonlocal invariant $\mathbf{H}[g](x)$ which has been
proposed to quantify inflationary back-reaction \cite{Tsamis:2013cka}.
In section \ref{1loop} we computed the one loop expectation value of
$\mathbf{H}[g](x)$, obtaining (\ref{H1exp}) for the contributions from that
part of $\mathbf{H}[g](x)$ which is linear in the graviton field, and 
(\ref{1loopH2}) for the contributions from the part of $\mathbf{H}[g](x)$
which is quadratic in the graviton field.

Section \ref{renorm} dealt with the crucial issue of renormalization.
The natural renormalization condition for the cosmological counterterm
$\delta \Lambda$ is to entirely cancel the trace of the 1PI 1-point 
function, which makes the 1-point contribution (\ref{H1exp}) vanish.
That leaves the divergent 2-point contribution (\ref{1loopH2}) uncontrolled.
These composite operator divergences require composite operator 
renormalization. We identified two candidate operators (\ref{mixingops})
which could be used to entirely cancel the 2-point contribution (\ref{1loopH2}),
at one loop order and on de Sitter background (and probably other homogeneous
and isotropic backgrounds). We have therefore confirmed the prediction that 
there is no back-reaction at one loop order, {\it and} we have a plausible 
conjecture for controlling ultraviolet divergences at any order and on 
general expanding spacetime backgrounds.

These are solid accomplishments which place the extension to two loop order
within reach. It is at this order that one expects secular back-reaction,
which cannot be absorbed by renormalization. Some of the additional work 
required for this project is mechanical:
\begin{itemize}
\item{Extend the expansion (\ref{Hexp}-\ref{H2}) of $\mathbf{H}[g](x)$ to
include terms with three and four powers of the graviton field;} 
\item{Re-do the old 2-loop computation of the 1PI 1-point function 
\cite{Tsamis:1996qm} using dimensional regularization;}
\item{Reduce the 2-loop 2-point contributions to either a single 4-point 
vertex with three propagators or two 3-point vertices with four propagators;}
\item{Reduce the 2-loop 3-point contributions to a single 3-point vertex
with three propagators; and}
\item{Reduce the 2-loop 4-point contributions to four propagators.}
\end{itemize}
Less mechanical is the task of including perturbative corrections to the
initial state wave function \cite{Kahya:2009sz}. It all seems doable now,
although the labor involved is certainly daunting.

Before closing we should comment on the possibility that the conjecture
of section \ref{renorm} might represent a new insight on how to renormalize 
nonlocal, composite operators. This has great significance for quantum
gravity because the only gauge invariant operators in that theory are
nonlocal. Recall that the problem with nonlocal composite operators is 
limiting the list of other operators with which they can mix. Our conjecture 
deals with the class of nonlocal composite operators that descend from a 
larger parent theory in which they are local, just as our expansion observable 
$\mathbf{H}[g](x)$ becomes local in the scalar + gravity theory (\ref{matterL}). 
We propose that the list of mixing operators be restricted to those which are 
local in the parent theory. The only other nonlocal composite operator whose 
renormalization is currently understood is the Wilson loop of non-Abelian 
gauge theory. They are multiplicatively renormalized \cite{Korchemsky:1987wg}, 
and that can indeed be viewed as a coupling constant renormalization in a 
parent theory which consists of a non-Abelian charged particle + Yang-Mills. 
It would be interesting to see if a similar result pertains for the quantum
gravitational analogue whose puzzling ultraviolet divergences are not currently 
understood \cite{Tsamis:1989yu}. {\it Node added in proof:} A recent study has 
partially confirmed this conjecture \cite{Frob:2017apy}.

\vskip 1cm

\centerline{\bf Acknowledgements}

This work was partially supported by Taiwan MOST grant
103-2112-M-006-001-MY3, by the European Union's Horizon 2020 Programme
under grant agreement 669288-SM-GRAV-ERC-2014-ADG; by NSF grant 
PHY-1506513; and by the Institute for Fundamental Theory at the 
University of Florida.

\section{Appendix A: General Scalar Propagator} \label{genpropdef}

Recall the general scalar kinetic operator $D_{\nu}$ which was defined
in expression (\ref{Dnu}). The spatial plane wave mode functions for
$D_{\nu}$ are,
\begin{equation}
u_{\nu}(\eta,k) = \sqrt{ \frac{\pi}{4 H a^{D-1}}} \, 
H^{(1)}_{\nu}(-k\eta) \; . \label{udef}
\end{equation}
Up to a possible infrared cutoff the associated propagator is,
\begin{eqnarray}
\lefteqn{ i\Delta_{\nu}(x;x') = \int \!\! \frac{d^{D-1}k}{(2\pi)^{D-1}} \,
e^{i \vec{k} \cdot (\vec{x} - \vec{x}')} \Biggl\{ \theta(\eta \!-\! \eta')
u_{\nu}(\eta,k) u^*_{\nu}(\eta',k) } \nonumber \\
& & \hspace{7cm} + \theta(\eta' \!-\! \eta) u^*_{\nu}(\eta,k) u_{\nu}(\eta',k)
\Biggr\} \; , \qquad \label{modesum}
\end{eqnarray}
Except for a handful of de Sitter breaking terms (for which see section 3 of
\cite{Miao:2010vs}) the result is,
\begin{eqnarray}
\lefteqn{i\Delta_{\nu}(x;x') = \frac{H^{D-2}}{(4\pi)^{\frac{D}2}} \Biggl\{
\Gamma\Bigl( \frac{D}2 \!-\! 1\Bigr) \Bigl( \frac{4}{y}\Bigr)^{\frac{D}2-1} -
\frac{\Gamma(\frac{D}2) \Gamma(1 \!-\! \frac{D}2)}{\Gamma(\frac12 \!+\! \nu) 
\Gamma(\frac12 \!-\! \nu)} \sum_{n=0}^{\infty} } \nonumber \\
& & \hspace{-.7cm} \times\! \Biggl[ 
\frac{\Gamma(\frac32 \!+\! \nu \!+\! n) \Gamma(\frac32 \!-\! \nu \!+\! n)}{
\Gamma(3 \!-\! \frac{D}2 \!+\! n) (n \!+\! 1)!} \Bigl( \frac{y}{4}
\Bigr)^{n-\frac{D}2+2} \!\!\!\!\!\!\!- \frac{\Gamma(\nu_A \!+\! \nu \!+\! n) 
\Gamma(\nu_A \!-\! \nu \!+\! n)}{\Gamma(\frac{D}2 \!+\! n) n!} 
\Bigl( \frac{y}{4}\Bigr)^n \Biggr] \Biggr\} . \label{propnu} \qquad
\end{eqnarray}
The special case of $\nu = \nu_A - N$ has great importance for us,
\begin{eqnarray}
\lefteqn{i\Delta_{\nu_A-N}(x;x') = \frac{H^{D-2}}{(4\pi)^{\frac{D}2}} \Biggl\{
\Gamma\Bigl( \frac{D}2 \!-\! 1\Bigr) \Bigl( \frac{4}{y}\Bigr)^{\frac{D}2-1} }
\nonumber \\
& & \hspace{2cm} + (-1)^N \sum_{n=0}^{\infty} \Biggl[ \frac{\Gamma(n \!-\! N 
\!+\! D \!-\! 1) \Gamma(n \!+\! N)}{\Gamma(n \!+\! \frac{D}2) \, n!} \Bigl( 
\frac{y}{4}\Bigr)^n \nonumber \\
& & \hspace{3.5cm} - \frac{\Gamma(n \!-\! N \!+\! \frac{D}2 \!+\! 1) 
\Gamma(n \!+\! N \!-\! \frac{D}2 \!+\! 2)}{\Gamma(n \!-\! \frac{D}2 \!+\! 3) 
\, (n \!+\! 1)!} \Bigl( \frac{y}{4} \Bigr)^{n-\frac{D}2+2} \Biggr] \Biggr\} . 
\label{propN} \qquad
\end{eqnarray}
Because all our results can be reduced to coincidence limits of differentiated
propagators it is worth pointing out that the potential ultraviolet divergences 
in expression (\ref{propN}) derive from the Gamma function $\Gamma(n-N+D-1)$
which multiplies the factor of $y^n$. Because $y(x;x')$ vanishes at coincidence,
nonzero results can only come from low powers of $y$. For certain integrated 
propagators such as $I_{ACA}(x;x')$ and $I_{AAA}(x;x')$ in expression 
(\ref{PhiPhiterms}) there can also be ultraviolet divergences from 
differentiating the multiplicative factor of $1/\Gamma(\frac12 - \nu)$ in 
expression (\ref{propnu}).

\section{Appendix B: Reflection Identities} \label{reflect}

In flat space background all components of the graviton and ghost
propagators are the same, and all depend only on the Lorentz invariant 
difference of the two points, $\eta_{\mu\nu} (x - x')^{\mu} (x - x')^{\nu}$.
It is therefore straightforward to reflect derivatives from one coordinate
of a propagator to the other, and from one side of an inverse differential
operator to the other,
\begin{equation}
{\rm Flat\ Space} \quad \Longrightarrow \quad \partial_{\mu} i\Delta(x;x') 
= -\partial_{\mu}' i\Delta(x;x') \quad , \quad \partial_{\mu} \frac1{\partial^2}
= \frac1{\partial^2} \partial_{\mu} \; . 
\end{equation} 
Expressions (\ref{gravprop}-\ref{ghostprop}) and (\ref{DeltaA}-\ref{DeltaC}) 
show that things are considerably more complicated on de Sitter background! 
However, it is still possible to reflect derivatives by extending some older 
relations \cite{Tsamis:1992zt}.

All propagators and inverse differential operators involve integrals of the
function (and its conjugate),
\begin{equation}
f_{\nu}\Bigl( \eta,\eta',\Delta \vec{x}\Bigr) \equiv u_{\nu}(\eta,k) 
u^*_{\nu}(\eta,k) e^{i \vec{k} \cdot (\vec{x} - \vec{x}')} \; .
\end{equation}
Using the Bessel function recursion relation $J_{\nu}'(z) \pm \frac{\nu}{z} 
J_{\nu}(z) = \pm J_{\nu \mp1}(z)$ we can reflect derivatives from one argument
to the other,
\begin{eqnarray}
\partial_i f_{\nu}\Bigl( \eta,\eta',\Delta \vec{x} \Bigr) & \!\!\!\!\!=\!\!\!\!\! 
& -\partial_i' f_{\nu}\Bigl( \eta,\eta',\Delta \vec{x} \Bigr) , \\
\Bigl[\partial_0 + (\nu_A \!-\! \nu) H a\Bigr] f_{\nu}\Bigl( \eta,\eta',\Delta 
\vec{x} \Bigr) & \!\!\!\!\!=\!\!\!\!\! &  -\Bigl[\partial_0' + (\nu_A \!+\! \nu 
\!-\! 1) H a' \Bigr] f_{\nu-1}\Bigl( \eta,\eta',\Delta \vec{x} \Bigr) , 
\qquad \\
\Bigl[\partial_0 + (\nu_A \!+\! \nu) H a\Bigr] f_{\nu}\Bigl( \eta,\eta',\Delta 
\vec{x} \Bigr) & \!\!\!\!\!=\!\!\!\!\! & -\Bigl[\partial_0' + (\nu_A \!-\! \nu 
\!-\! 1) H a'\Bigr] f_{\nu+1}\Bigl( \eta,\eta',\Delta \vec{x} \Bigr) . \qquad 
\end{eqnarray} 
Applying these identities to the propagator implies,
\begin{eqnarray}
\partial_i i\Delta_{\nu}\Bigl(x;x') & \!\!\!=\!\!\! & -\partial_i' 
i\Delta_{\nu}(x;x') \; , \label{diprop} \\
\Bigl[\partial_0 + (\nu_A \!-\! \nu) H a\Bigr] i\Delta_{\nu}(x;x')
& \!\!\!=\!\!\! &  -\Bigl[\partial_0' + (\nu_A \!+\! \nu \!-\! 1) H a' \Bigr] 
i\Delta_{\nu-1}(x;x') \; , \qquad \label{d0propdown} \\
\Bigl[\partial_0 + (\nu_A \!+\! \nu) H a\Bigr] i\Delta_{\nu}(x;x')
& \!\!\!=\!\!\! & -\Bigl[\partial_0' + (\nu_A \!-\! \nu \!-\! 1) H a'\Bigr] 
i\Delta_{\nu+1}(x;x') \; . \qquad \label{d0propup} 
\end{eqnarray}
The analogous relations for inverse differential operators are,
\begin{eqnarray}
\frac1{D_{\nu}} \, \partial_i & = & \partial_i \frac1{D_{\nu}} \; , 
\label{diDinv} \\
\frac1{D_{\nu}} \, \Bigl[\partial_0 - (\nu_A \!-\! \nu) H a\Bigr] & = &  
\Bigl[\partial_0 + (\nu_A \!+\! \nu \!-\! 1) H a \Bigr] \frac1{D_{\nu-1}} \; , 
\qquad \label{d0Dinvdown} \\
\frac1{D_{\nu}} \, \Bigl[\partial_0 - (\nu_A \!+\! \nu) H a\Bigr] 
& = & \Bigl[\partial_0 + (\nu_A \!-\! \nu \!-\! 1) H a\Bigr] \frac1{D_{\nu+1}}
 \; . \qquad \label{d0Dinvup}
\end{eqnarray}

\section{Appendix C: Integrated Propagators} \label{intprop}

The integrated propagators $I_{\alpha\beta}$ and $I_{\alpha\beta\gamma}$ 
of expression (\ref{Iints}) are symmetric,
\begin{equation}
I_{\alpha\beta}(x;x') = I_{\beta\alpha}(x;x') \qquad , \qquad 
I_{\alpha\beta\gamma}(x;x') = I_{\beta\alpha\gamma}(x;x') =
I_{\alpha\gamma\beta}(x;x') \; . \label{Isyms}
\end{equation}
They can also be generalized to any number of integrations,
\begin{equation}
I_{\alpha\beta\cdots\psi\omega}(x;x') = \frac1{D_{\alpha}} \frac1{D_{\beta}}
\cdots \frac1{D_{\psi}} \, i\Delta_{\omega}(x;x') \; . \label{genI}
\end{equation}
By counting inverse derivatives one can easily infer the leading behavior
of these integrated propagators near coincidence,
\begin{equation}
I_{\alpha_1 \cdots \alpha_{n}}(x;x') \sim \Delta x^{2n-4} \ln(\Delta x^2)
\qquad , \qquad \Delta x^2 \equiv \eta_{\mu\nu} (x \!-\! x')^{\mu} 
(x \!-\! x')^{\nu} \; . \label{Icoinc}
\end{equation}
So the integrated propagators $\nabla^2 I_{AAA}(x;x')$ and $\nabla^2
I_{ACA}(x;x')$ in expression (\ref{PhiPhiterms}) are only logarithmically 
divergent at coincidence.  

The $(n+1)$-th integrated propagator can be written simply in terms of 
differences of the $n$-th integrated propagators \cite{Miao:2010vs},
\begin{eqnarray}
I_{\alpha\beta}(x;x') & = & \frac{i\Delta_{\alpha}(x;x') \!-\! 
i\Delta_{\beta}(x;x')}{(\beta^2 \!-\! \alpha^2) H^2} \; , \label{2subs} \\
I_{\alpha\beta\gamma}(x;x') & = & \frac{I_{\alpha\gamma}(x;x') \!-\! 
I_{\beta\gamma}(x;x')}{(\beta^2 \!-\! \alpha^2) H^2} \; , \label{3subs}
\end{eqnarray}
and so on. It follows that the coincidence limits of derivatives of
integrated propagators which are given in expressions (\ref{hPhiterms}) 
and (\ref{PhiPhiterms}) are really coincidence limits of differences of
differentiated propagators. In dimensional regularization these 
coincidence limits come entirely from the first few $y^n$ terms. For 
example, the contributions from $I_{AC} = -(i\Delta_A - 
i\Delta_C)/[2(D-3) H^2]$ in expression (\ref{hPhiterms}) derive from 
just the de Sitter breaking factor of $\ln(a a')$ and the $y^1$ terms of 
the two propagators (\ref{DeltaA}) and (\ref{DeltaC}),
\begin{eqnarray}
Ha \partial_0 I_{AC} \Bigr\vert_{x' = x} & \!\!\!=\!\!\! & 
\frac{-H^{D-4}}{2 (D \!-\!3) (4\pi)^{\frac{D}2}} 
\frac{\Gamma(D \!-\! 1)}{\Gamma(\frac{D}2)} \Biggl\{ 1 + 0 \Biggr\} 
\!\times\! H^2 a^2 \; , \label{d0IAC} \\
\partial_0^2 I_{AC} \Bigr\vert_{x' = x} & \!\!\!=\!\!\! & 
\frac{-H^{D-4}}{2 (D \!-\!3) (4\pi)^{\frac{D}2}} 
\frac{\Gamma(D \!-\! 1)}{\Gamma(\frac{D}2)} \Biggl\{ 1 - \Bigl(
\frac{D \!-\! 3}{D \!-\! 2}\Bigr) \Biggr\} \!\times\! H^2 a^2 \; , 
\label{d02IAC} \\
\nabla^2 I_{AC} \Bigr\vert_{x' = x} & \!\!\!=\!\!\! & 
\frac{-H^{D-4}}{2 (D \!-\!3) (4\pi)^{\frac{D}2}} 
\frac{\Gamma(D \!-\! 1)}{\Gamma(\frac{D}2)} \Biggl\{ 0 + 
\frac{(D \!-\! 1) (D \!-\! 3)}{D \!-\! 2} \Biggr\} \!\times\! H^2 a^2 \; .
\qquad \label{nablaIAC}
\end{eqnarray}

Repeated subscripts, such as those in $I_{ACA}$ and $I_{AAA}$ of expression
(\ref{PhiPhiterms}), follow from expressions (\ref{2subs}-\ref{3subs}) by
differentiation with respect to the subscript,
\begin{eqnarray}
I_{\alpha\alpha}(x;x') & = & -\frac1{2\alpha H^2} \, 
\frac{\partial i\Delta_{\nu}(x;x')}{\partial \nu} \Bigl\vert_{\nu = \alpha}
\; , \qquad \label{Iaa} \\
I_{\alpha\alpha\beta}(x;x') & = & -\frac{I_{\alpha\beta}(x;x')}{(\beta^2 \!-\!
\alpha^2) H^2} - \frac1{2 \alpha (\beta^2 \!-\! \alpha^2) H^4} \,
\frac{\partial i\Delta_{\nu}(x;x')}{\partial \nu} \Bigl\vert_{\nu = \alpha} 
\; , \qquad \label{Iaab} \\
I_{\alpha\alpha\alpha}(x;x') & = & \frac{I_{\alpha\alpha}(x;x')}{4 \alpha^2 H^2}
+ \frac1{8 \alpha^2 H^4} \, \frac{\partial^2 i\Delta_{\nu}(x;x')}{\partial \nu^2}
\Bigl\vert_{\nu = \alpha} \; . \qquad \label{Iaaa}
\end{eqnarray}
All the triple subscript integrated propagators in expression (\ref{PhiPhiterms})
involve $\frac{\nabla^2}{a^2}$. This has the effect of eliminating the purely 
time dependent, de Sitter breaking terms. In view of relations 
(\ref{Iaab}-\ref{Iaaa}) the result we need is,
\begin{eqnarray}
\lefteqn{\frac{\nabla^2}{a^2} \, i\Delta_{\nu}(x;x') \Bigl\vert_{x'=x} }
\nonumber \\
& & \hspace{.5cm} = 
\frac{H^{D-2}}{(4\pi)^{\frac{D}2}} \frac{\Gamma(\frac{D}2) \Gamma(1 \!-\! 
\frac{D}2)}{\Gamma(\frac12 \!+\! \nu) \Gamma(\frac12 \!-\! \nu)} 
\frac{\Gamma(\nu_A \!+\! \nu \!+\! 1) \Gamma(\nu_A \!-\! \nu \!+\! 1)}{
\Gamma(\frac{D}2 \!+\! 1)} \!\times\! \frac12 (D\!-\!1) H^2 \; . \qquad 
\label{keyeqn}
\end{eqnarray}
Differentiating expression (\ref{keyeqn}) with respect to $\nu$ and
setting $\nu = \nu_A$ gives,
\begin{eqnarray}
\lefteqn{\frac{\nabla^2}{a^2} \frac{\partial i\Delta_{\nu}}{\partial \nu} 
\Bigl\vert_{x' = x \atop \nu = \nu_A} = \frac{H^D}{(4\pi)^{\frac{D}2}} 
\frac{(D \!-\! 1) \Gamma(D)}{D \Gamma(\frac{D}2)} \Biggl\{\psi(D) \!-\! 
\psi(1) \!-\! \psi\Bigl(\frac{D}2\Bigr) \!+\! \psi\Bigl(1 \!-\! \frac{D}2
\Bigr) \Biggr\} , } \label{dnu1} \\
\lefteqn{\frac{\nabla^2}{a^2} \frac{\partial^2 i\Delta_{\nu}}{\partial \nu^2} 
\Bigl\vert_{x' = x \atop \nu = \nu_A} = \frac{H^D}{(4\pi)^{\frac{D}2}} 
\frac{(D \!-\!1) \Gamma(D)}{\Gamma(D \frac{D}2)} \Biggl\{\psi'(D) \!+\! 
\psi'(1) \!-\! \psi'\Bigl(\frac{D}2\Bigr) \!-\! \psi'\Bigl(1 \!-\! \frac{D}2
\Bigr) } \nonumber \\
& & \hspace{5cm} + \Biggl[\psi(D) \!-\! 
\psi(1) \!-\! \psi\Bigl(\frac{D}2\Bigr) \!+\! \psi\Bigl(1 \!-\! \frac{D}2
\Bigr)\Biggr]^2 \Biggr\} . \qquad \label{dnu2}
\end{eqnarray}
Retaining the full $D$-dependence becomes extremely tedious so we report only
the divergent contributions,
\begin{eqnarray}
\frac{\nabla^2}{a^2} \frac{\partial i\Delta_{\nu}}{\partial \nu} 
\Bigl\vert_{x' = x \atop \nu = \nu_A} & = & \frac{H^D}{(4\pi)^{\frac{D}2}} 
\frac{(\Gamma(D\!-\!1)}{\Gamma(\frac{D}2)} \Biggl\{ \frac{\frac{9}{2}}{D \!-\! 4}
+ O\Bigl( (D \!-\! 4)^0\Bigr) \Biggr\} , \label{dnu1simp} \\
\frac{\nabla^2}{a^2} \frac{\partial^2 i\Delta_{\nu}}{\partial \nu^2} 
\Bigl\vert_{x' = x \atop \nu = \nu_A} & = & \frac{H^D}{(4\pi)^{\frac{D}2}} 
\frac{(\Gamma(D\!-\!1)}{\Gamma(\frac{D}2)} \Biggl\{ \frac{\frac{33}{2}}{D \!-\! 4}
+ O\Bigl( (D \!-\! 4)^0\Bigr) \Biggr\} . \qquad \label{dnu2simp}
\end{eqnarray}
Employing relations (\ref{dnu1simp}-\ref{dnu2simp}) and expressions 
(\ref{Iaab}-\ref{Iaaa}) gives,
\begin{eqnarray}
\frac{\nabla^2}{a^2} I_{ACA}(x;x') \Bigl\vert_{x'=x} & = & 
\frac{H^{D-4}}{(4\pi)^{\frac{D}2}} \frac{(\Gamma(D\!-\!1)}{\Gamma(\frac{D}2)} 
\Biggl\{ \frac{\frac{3}{4}}{D \!-\! 4} + O\Bigl( (D \!-\! 4)^0\Bigr) \Biggr\} , 
\label{IAAC} \\
\frac{\nabla^2}{a^2} I_{AAA}(x;x') \Bigl\vert_{x'=x} & = & 
\frac{H^{D-4}}{(4\pi)^{\frac{D}2}} \frac{(\Gamma(D\!-\!1)}{\Gamma(\frac{D}2)} 
\Biggl\{ \frac{\frac{3}{4}}{D \!-\! 4} + O\Bigl( (D \!-\! 4)^0\Bigr) \Biggr\} , 
\label{IAAA}
\end{eqnarray}

The inverse factors of $a$ and $a'$ in the integrated propagators of
expressions (\ref{Jints}-\ref{JKints}) require additional labor. By inserting 
unity in the form $1 = D_{\alpha} \times \frac1{D_{\alpha}}$, partially 
integrating, and then using the reflection identities of Appendix B we can 
expand $J_{\mu\nu}(x;x')$ in terms of ever more highly integrated propagators,
\begin{eqnarray}
J_{\mu\nu}(x;x') & = & \sum_{K=0}^{\infty} K! (-2H)^K \Biggl[ 
\partial_0 + \Bigl(\nu_A \!+\! \mu \!-\! 1\Bigr) H a\Biggr] \cdots \nonumber \\
& & \hspace{1cm} \cdots \Biggl[ \partial_0 + \Bigl(\nu_A \!+\! \mu \!-\! K\Bigr) 
H a\Biggr] \Biggl\{ \frac{ I_{\nu \, \mu-1 \cdots \mu-2K-1}(x;x')}{a^{K+1}} 
\Biggr\} , \qquad \label{J1} \\
& = & \sum_{K=0}^{\infty} K! (-2H)^K \Biggl[ \partial'_0 + \Bigl(\nu_A \!+\! 
\nu \!-\! 1\Bigr) H a'\Biggr] \cdots \nonumber \\
& & \hspace{1cm} \cdots \Biggl[ \partial'_0 + \Bigl(\nu_A \!+\! \nu \!-\! K\Bigr) 
H a'\Biggr] \Biggl\{ \frac{ I_{\mu \, \nu-1 \cdots \nu-2K-1}(x;x')}{{a'}^{K+1}} 
\Biggr\} . \qquad \label{J2}
\end{eqnarray}
Because either argument can be chosen for the expansion it is possible to avoid 
repeated indices in the expansions for $J_{AD}$ and $J_{AB}$, for example,
\begin{eqnarray}
\lefteqn{J_{AD}(x';x) = \frac{I_{AE}}{a} - 2 H \Bigl[ \partial_0 + 
(D \!-\! 5) H a\Bigr] \frac{I_{AEG}}{a^2} } \nonumber \\
& & \hspace{-.5cm} + 8 H^2 \Bigl[ \partial_0 + (D \!-\! 5) H a\Bigr] 
\Bigl[\partial_0 + (D \!-\! 6) H a\Bigr] \frac{I_{AEGI}}{a^3} \nonumber \\
& & \hspace{-.5cm} - 48 H^3 \Bigl[ \partial_0 \!+\! (D \!-\! 5) H a\Bigr] 
\Bigl[\partial_0 \!+\! (D \!-\! 6) H a\Bigr] \Bigl[\partial_0 \!+\! 
(D \!-\! 7) H a\Bigr] \frac{I_{AEGIK}}{a^4} + \dots \qquad \label{JADexp}
\end{eqnarray}

\begin{table}
\setlength{\tabcolsep}{8pt}
\def\arraystretch{1.5}
\centering
\begin{tabular}{|@{\hskip 1mm }c@{\hskip 1mm }|@{\hskip 1mm }c@{\hskip 1mm }||c|c|} 
\hline
$K$ & $n$ & Terms for $J_{AD}$ & Terms for $J_{AB}$ \\
\hline\hline
0 & 0 & $0$ & $0$ \\
\hline
0 & 1 & $0$ & $-\frac{9}{4}$ \\
\hline
1 & 0 & $0$ & $0$ \\
\hline
1 & 1 & $0$ & $0$ \\
\hline
1 & 2 & $-5$ & $-2$ \\
\hline
2 & 0 & $0$ & $0$ \\
\hline
2 & 1 & $0$ & $0$ \\
\hline
2 & 2 & $-144$ & $-6$ \\
\hline
3 & 0 & $0$ & $0$ \\
\hline
3 & 1 & $0$ & $0$ \\
\hline
3 & 2 & $-7460$ & $-216$ \\
\hline
3 & 3 & $756$ & $324$ \\
\hline
\end{tabular}
\caption{The results of acting the external derivatives of expressions
(\ref{JAD}-\ref{JAB}) on the relevant internal derivatives (\ref{intders}),
and then taking the coincidence limit for $D=4$. For $J_{AD}$ we used $N=3$,
with $N=1$ for $J_{AB}$. \label{JADAB}}
\end{table}

The integrated propagators $J_{AD}(x';x)$ and $J_{AB}(x';x)$ appear in 
expression (\ref{hPhiterms}) with certain external derivatives,
\begin{eqnarray}
\frac1{a^2} \Biggl[ \partial_0^2 + (D\!-\!4) H a \partial_0 - 
\frac{2 \nabla^2}{D \!-\! 1}\Biggr] \Bigl[\partial_0 + (D\!-\!4) Ha\Bigr]
\frac{J_{AD}(x';x)}{D \!-\! 3} \Bigl\vert_{x'=x} \; , \label{JAD} \\
-\frac1{a^2} \Biggl[ \frac{[\partial_0^2 \!-\! Ha\partial_0 \!-\! \nabla^2]
[\partial_0 + (D \!-\! 2) Ha]}{D \!-\! 3} + \frac{\nabla^2 (\partial_0 \!-\!
H a)}{D \!-\! 1} \Biggr] J_{AB}(x';x) \Bigl\vert_{x'=x} \; . \label{JAB} 
\end{eqnarray}
In view of relation (\ref{Icoinc}) the only possible divergences can arise 
for $K \leq 3$. Because the various integrated propagators $I_{\alpha \cdots
\beta}$ involve differences of propagators (\ref{propN}), the result for 
(\ref{JAD}-\ref{JAB}) derives from acting the appropriate external derivatives 
on expressions of the form,
\begin{equation}
\frac{K! (-2H)^K}{H^{2K-2}} \Bigl[\partial_0 + (D\!-\!2\!-\!N) Ha\Bigr] \cdots
\Bigl[\partial_0 + (D\!-\! 2 \!-\! N \!-\! K) Ha \Bigr] \Bigl( \frac{y}{4}
\Bigr)^n \; . \label{intders}
\end{equation}
Table \ref{JADAB} gives the results of doing this for the few values of $K$
and $n$ which are required. These factors are then multiplied by the ratios 
of Gamma functions from (\ref{propN}) for each value of $n$ and $N$ ($N=3$ for
$J_{AD}$ and $N=1$ for $J_{AB}$), and finally divided by the numerical factors 
implied by expressions (\ref{2subs}-\ref{3subs}). Putting everything together 
gives,
\begin{eqnarray}
\Bigl(\ref{JAD}\Bigr) & = & \frac{H^{D-1}}{(4\pi)^{\frac{D}2}} 
\frac{\Gamma(D \!-\! 1)}{\Gamma(\frac{D}2)} \Biggl\{ \frac{\frac{15227}{780}}{
D \!-\! 4} + O\Bigl( (D \!-\! 4)\Bigr) \Biggr\} , \label{JADfinal} \\
\Bigl(\ref{JAB}\Bigr) & = & \frac{H^{D-1}}{(4\pi)^{\frac{D}2}} 
\frac{\Gamma(D \!-\! 1)}{\Gamma(\frac{D}2)} \Biggl\{ \frac{-\frac{42443}{924}}{
D \!-\! 4} + O\Bigl( (D \!-\! 4)\Bigr) \Biggr\} . \label{JABfinal}
\end{eqnarray}
Because the $I_{AC}$ contributions (\ref{d0IAC}-\ref{nablaIAC}) are all finite,
the divergent part of (\ref{hPhiterms}) comes from adding (\ref{JADfinal}) to
(\ref{JABfinal}) and multiplying by $\kappa^2$.

Doubly integrated propagators with inverse factors of $a$ and $a'$ require
a separate treatment. One first writes them as a single inverse differential
operator acting on a singly integrated propagator, then the singly integrated
propagator is expanded according to expressions (\ref{2subs}) or 
(\ref{J1}-\ref{J2}). For example, consider $K_{ACB}(x;x')$ from expression
(\ref{PhiPhiterms}),
\begin{eqnarray}
\lefteqn{K_{ACB}(x;x') = \frac1{D_B'} \Bigl[ \frac{I_{AC}(x;x')}{a'} \Bigr] =
\frac{J_{BC}(x';x) \!-\! J_{BA}(x';x)}{2 (D\!-\!3) H^2} \; , } \\
& & \hspace{1cm} = \frac1{2 (D\!-\!3) H^2} \Biggl\{ \frac{I_{BD}}{a} - 2H
\Bigl[\partial_0 + (D\!-\! 4) Ha\Bigr] \frac{I_{BDF}}{a^2} + \dots 
\nonumber \\
& & \hspace{4cm} - \frac{I_{CA}}{a'} + 2 H \Bigl[\partial_0' + (D\!-\!3) Ha'
\Bigr] \frac{I_{CEA}}{{a'}^2} - \dots \Biggr\} . \qquad \label{Kexp}
\end{eqnarray}
Because $K_{ACB}(x;x')$ is only differentiated three times in expression
(\ref{PhiPhiterms}), we do not need to go any higher than the terms shown
in (\ref{Kexp}). The three $J$ integrals in expression (\ref{PhiPhiterms})
are differentiated four times so they must be expanded to one higher order. 
Our final results for the $J$ and $K$ integrals in expression
(\ref{PhiPhiterms}) are,
\begin{eqnarray}
\lefteqn{ \lim_{x'=x} \frac{\nabla^2}{a^2} \Bigl[ \partial_0' + (D\!-\!2) 
Ha' \Bigr] H K_{AAB}(x;x') } \nonumber \\
& & \hspace{4.5cm} = \frac{H^{D-2}}{(4\pi)^{\frac{D}2}} \frac{\Gamma(D \!-\! 1)}{
\Gamma(\frac{D}2)} \Biggl\{ \frac{\frac74}{D \!-\! 4} + O\Bigl( (D \!-\! 4)\Bigr) 
\Biggr\} , \label{KAAB} \qquad \\
\lefteqn{ \lim_{x'=x} \frac{\nabla^2}{a^2} \Bigl[ \partial_0' + (D\!-\!2) 
Ha' \Bigr] H K_{ACB}(x;x') } \nonumber \\
& & \hspace{4.5cm} = \frac{H^{D-2}}{(4\pi)^{\frac{D}2}} \frac{\Gamma(D \!-\! 1)}{
\Gamma(\frac{D}2)} \Biggl\{ \frac{\frac{3}{320}}{D \!-\! 4} + O\Bigl( 
(D \!-\! 4)\Bigr) \Biggr\} , \label{KACB} \qquad \\
\lefteqn{ \lim_{x'=x} \frac{\nabla^4}{a^2} J_{ABA}(x;x') } \nonumber \\
& & \hspace{4.5cm} = \frac{H^{D-2}}{(4\pi)^{\frac{D}2}} \frac{\Gamma(D \!-\! 1)}{
\Gamma(\frac{D}2)} \Biggl\{ \frac{\frac{145}{16}}{D \!-\! 4} + O\Bigl( (D \!-\! 4)
\Bigr) \Biggr\} , \label{JABA} \qquad \\
\lefteqn{ \lim_{x'=x} \frac{\nabla^2}{a^2} \Bigl[\partial_0 + (D\!-\!2) Ha\Bigr]
\Bigl[ \partial_0' + (D\!-\!2) Ha' \Bigr] J_{BAB}(x;x') } \nonumber \\
& & \hspace{4.5cm} = \frac{H^{D-2}}{(4\pi)^{\frac{D}2}} \frac{\Gamma(D \!-\! 1)}{
\Gamma(\frac{D}2)} \Biggl\{ \frac{\frac{51}{16}}{D \!-\! 4} + O\Bigl( (D \!-\! 4)
\Bigr) \Biggr\} , \label{JBAB} \qquad \\
\lefteqn{ \lim_{x'=x} \frac{\nabla^2}{a^2} \Bigl[\partial_0 + (D\!-\!2) Ha\Bigr]
\Bigl[ \partial_0' + (D\!-\!2) Ha' \Bigr] J_{BCB}(x;x') } \nonumber \\
& & \hspace{4.5cm} = \frac{H^{D-2}}{(4\pi)^{\frac{D}2}} \frac{\Gamma(D \!-\! 1)}{
\Gamma(\frac{D}2)} \Biggl\{ \frac{\frac{163}{120}}{D \!-\! 4} + O\Bigl( (D \!-\! 4)
\Bigr) \Biggr\} . \label{JBCB} \qquad 
\end{eqnarray}


\begin{thebibliography}{99}

\bibitem{Weinberg:1988cp} 
  S.~Weinberg,
  Rev.\ Mod.\ Phys.\  {\bf 61}, 1 (1989).
  doi:10.1103/RevModPhys.61.1

\bibitem{Carroll:2000fy} 
  S.~M.~Carroll,
  Living Rev.\ Rel.\  {\bf 4}, 1 (2001)
  doi:10.12942/lrr-2001-1
  [astro-ph/0004075].

\bibitem{Mukhanov:2005sc} 
  V.~Mukhanov,
  ``Physical Foundations of Cosmology,''
  Cambridge UK: Cambridge University Press (2005) 421 P.

\bibitem{Tsamis:1996qq} 
  N.~C.~Tsamis and R.~P.~Woodard,
  Nucl.\ Phys.\ B {\bf 474}, 235 (1996)
  doi:10.1016/0550-3213(96)00246-5
  [hep-ph/9602315].

\bibitem{Tsamis:2011ep} 
  N.~C.~Tsamis and R.~P.~Woodard,
  Int.\ J.\ Mod.\ Phys.\ D {\bf 20}, 2847 (2011)
  doi:10.1142/S0218271811020652
  [arXiv:1103.5134 [gr-qc]].
  
\bibitem{Parker:1969au} 
  L.~Parker,
  Phys.\ Rev.\  {\bf 183}, 1057 (1969).
  doi:10.1103/PhysRev.183.1057

\bibitem{Grishchuk:1974ny} 
  L.~P.~Grishchuk,
  Sov.\ Phys.\ JETP {\bf 40}, 409 (1975)
  [Zh.\ Eksp.\ Teor.\ Fiz.\  {\bf 67}, 825 (1974)].

\bibitem{Starobinsky:1979ty} 
  A.~A.~Starobinsky,
  JETP Lett.\  {\bf 30}, 682 (1979)
  [Pisma Zh.\ Eksp.\ Teor.\ Fiz.\  {\bf 30}, 719 (1979)].

\bibitem{Mukhanov:1981xt} 
  V.~F.~Mukhanov and G.~V.~Chibisov,
  JETP Lett.\  {\bf 33}, 532 (1981)
  [Pisma Zh.\ Eksp.\ Teor.\ Fiz.\  {\bf 33}, 549 (1981)].

\bibitem{Polyakov:1982ug} 
  A.~M.~Polyakov,
  Sov.\ Phys.\ Usp.\  {\bf 25}, 187 (1982)
  [Usp.\ Fiz.\ Nauk {\bf 136}, 538 (1982)].
  doi:10.1070/PU1982v025n03ABEH004529
             
\bibitem{Myhrvold:1983hx} 
  N.~P.~Myhrvold,
  Phys.\ Rev.\ D {\bf 28}, 2439 (1983).
  doi:10.1103/PhysRevD.28.2439

\bibitem{Ford:1984hs} 
  L.~H.~Ford,
  Phys.\ Rev.\ D {\bf 31}, 710 (1985).
  doi:10.1103/PhysRevD.31.710
  
\bibitem{Mottola:1984ar} 
  E.~Mottola,
  Phys.\ Rev.\ D {\bf 31}, 754 (1985).
  doi:10.1103/PhysRevD.31.754

\bibitem{Antoniadis:1985pj} 
  I.~Antoniadis, J.~Iliopoulos and T.~N.~Tomaras,
  Phys.\ Rev.\ Lett.\  {\bf 56}, 1319 (1986).
  doi:10.1103/PhysRevLett.56.1319

\bibitem{Mazur:1986et} 
  P.~Mazur and E.~Mottola,
  Nucl.\ Phys.\ B {\bf 278}, 694 (1986).
  doi:10.1016/0550-3213(86)90058-1

\bibitem{Antoniadis:1991fa} 
  I.~Antoniadis and E.~Mottola,
  Phys.\ Rev.\ D {\bf 45}, 2013 (1992).
  doi:10.1103/PhysRevD.45.2013
  
\bibitem{Tsamis:1992sx} 
  N.~C.~Tsamis and R.~P.~Woodard,
  Phys.\ Lett.\ B {\bf 301}, 351 (1993).
  doi:10.1016/0370-2693(93)91162-G
  
\bibitem{Krotov:2010ma} 
  D.~Krotov and A.~M.~Polyakov,
  Nucl.\ Phys.\ B {\bf 849}, 410 (2011)
  doi:10.1016/j.nuclphysb.2011.03.025
  [arXiv:1012.2107 [hep-th]].

\bibitem{Akhmedov:2012pa} 
  E.~T.~Akhmedov and P.~Burda,
  Phys.\ Rev.\ D {\bf 86}, 044031 (2012)
  doi:10.1103/PhysRevD.86.044031
  [arXiv:1202.1202 [hep-th]].

\bibitem{Polyakov:2012uc} 
  A.~M.~Polyakov,
  arXiv:1209.4135 [hep-th].
  
\bibitem{Anderson:2013zia} 
  P.~R.~Anderson and E.~Mottola,
  Phys.\ Rev.\ D {\bf 89}, 104039 (2014)
  doi:10.1103/PhysRevD.89.104039
  [arXiv:1310.1963 [gr-qc]].

\bibitem{Mukhanov:1996ak} 
  V.~F.~Mukhanov, L.~R.~W.~Abramo and R.~H.~Brandenberger,
  Phys.\ Rev.\ Lett.\  {\bf 78}, 1624 (1997)
  doi:10.1103/PhysRevLett.78.1624
  [gr-qc/9609026].

\bibitem{Abramo:1997hu} 
  L.~R.~W.~Abramo, R.~H.~Brandenberger and V.~F.~Mukhanov,
  Phys.\ Rev.\ D {\bf 56}, 3248 (1997)
  doi:10.1103/PhysRevD.56.3248
  [gr-qc/9704037].

\bibitem{Unruh:1998ic} 
  W.~Unruh,
  astro-ph/9802323.

\bibitem{Abramo:1998hi} 
  L.~R.~W.~Abramo and R.~P.~Woodard,
  Phys.\ Rev.\ D {\bf 60}, 044010 (1999)
  doi:10.1103/PhysRevD.60.044010
  [astro-ph/9811430].

\bibitem{Abramo:1998hj} 
  L.~R.~W.~Abramo and R.~P.~Woodard,
  Phys.\ Rev.\ D {\bf 60}, 044011 (1999)
  doi:10.1103/PhysRevD.60.044011
  [astro-ph/9811431].
           
\bibitem{Abramo:2001db} 
  L.~R.~Abramo and R.~P.~Woodard,
  Phys.\ Rev.\ D {\bf 65}, 043507 (2002)
  doi:10.1103/PhysRevD.65.043507
  [astro-ph/0109271].
  
\bibitem{Abramo:2001dc} 
  L.~R.~Abramo and R.~P.~Woodard,
  Phys.\ Rev.\ D {\bf 65}, 063515 (2002)
  doi:10.1103/PhysRevD.65.063515
  [astro-ph/0109272].
  
\bibitem{Geshnizjani:2002wp} 
  G.~Geshnizjani and R.~Brandenberger,
  Phys.\ Rev.\ D {\bf 66}, 123507 (2002)
  doi:10.1103/PhysRevD.66.123507
  [gr-qc/0204074].

\bibitem{Geshnizjani:2003cn} 
  G.~Geshnizjani and R.~Brandenberger,
  JCAP {\bf 0504}, 006 (2005)
  doi:10.1088/1475-7516/2005/04/006
  [hep-th/0310265].

\bibitem{Marozzi:2011zb} 
  G.~Marozzi and G.~P.~Vacca,
  Class.\ Quant.\ Grav.\  {\bf 29}, 115007 (2012)
  doi:10.1088/0264-9381/29/11/115007
  [arXiv:1108.1363 [gr-qc]].

\bibitem{Marozzi:2012tp} 
  G.~Marozzi, G.~P.~Vacca and R.~H.~Brandenberger,
  JCAP {\bf 1302}, 027 (2013)
  doi:10.1088/1475-7516/2013/02/027
  [arXiv:1212.6029 [hep-th]].

\bibitem{Tsamis:2013cka} 
  N.~C.~Tsamis and R.~P.~Woodard,
  Phys.\ Rev.\ D {\bf 88}, no. 4, 044040 (2013)
  doi:10.1103/PhysRevD.88.044040
  [arXiv:1306.6441 [gr-qc]].

\bibitem{Tsamis:1992xa} 
  N.~C.~Tsamis and R.~P.~Woodard,
  Commun.\ Math.\ Phys.\  {\bf 162}, 217 (1994).
  doi:10.1007/BF02102015

\bibitem{Woodard:2004ut} 
  R.~P.~Woodard,
  gr-qc/0408002.

\bibitem{Bogoliubov:1957gp} 
  N.~N.~Bogoliubov and O.~S.~Parasiuk,
  Acta Math.\  {\bf 97}, 227 (1957).
  doi:10.1007/BF02392399

\bibitem{Hepp:1966eg} 
  K.~Hepp,
  Commun.\ Math.\ Phys.\  {\bf 2}, 301 (1966).
  doi:10.1007/BF01773358

\bibitem{Zimmermann:1968mu} 
  W.~Zimmermann,
  Commun.\ Math.\ Phys.\  {\bf 11}, 1 (1968).
  doi:10.1007/BF01654298

\bibitem{Zimmermann:1969jj} 
  W.~Zimmermann,
  Commun.\ Math.\ Phys.\  {\bf 15}, 208 (1969)
  [Lect.\ Notes Phys.\  {\bf 558}, 217 (2000)].
  doi:10.1007/BF01645676
              
\bibitem{Itzykson:1980rh} 
  C.~Itzykson and J.~B.~Zuber,
  ``Quantum Field Theory,''
  New York, Usa: Mcgraw-Hill (1980) 705 P.

\bibitem{Weinberg:1996kr} 
  S.~Weinberg,
  ``The quantum theory of fields. Vol. 2: Modern applications,''
  Cambridge, UK: Cambridge University Press (1996) 489 P.
  
\bibitem{Korchemsky:1987wg} 
  G.~P.~Korchemsky and A.~V.~Radyushkin,
  Nucl.\ Phys.\ B {\bf 283}, 342 (1987).
  doi:10.1016/0550-3213(87)90277-X

\bibitem{Onemli:2002hr} 
  V.~K.~Onemli and R.~P.~Woodard,
  Class.\ Quant.\ Grav.\  {\bf 19}, 4607 (2002)
  doi:10.1088/0264-9381/19/17/311
  [gr-qc/0204065].

\bibitem{Onemli:2004mb} 
  V.~K.~Onemli and R.~P.~Woodard,
  Phys.\ Rev.\ D {\bf 70}, 107301 (2004)
  doi:10.1103/PhysRevD.70.107301
  [gr-qc/0406098].

\bibitem{Tsamis:2005je} 
  N.~C.~Tsamis and R.~P.~Woodard,
  Annals Phys.\  {\bf 321}, 875 (2006)
  doi:10.1016/j.aop.2005.08.004
  [gr-qc/0506056].

\bibitem{Tsamis:1996qm} 
  N.~C.~Tsamis and R.~P.~Woodard,
  Annals Phys.\  {\bf 253}, 1 (1997)
  doi:10.1006/aphy.1997.5613
  [hep-ph/9602316].
  
\bibitem{Iliopoulos:1998wq} 
  J.~Iliopoulos, T.~N.~Tomaras, N.~C.~Tsamis and R.~P.~Woodard,
  Nucl.\ Phys.\ B {\bf 534}, 419 (1998)
  doi:10.1016/S0550-3213(98)00528-8
  [gr-qc/9801028].
    
\bibitem{Kahya:2009sz} 
  E.~O.~Kahya, V.~K.~Onemli and R.~P.~Woodard,
  Phys.\ Rev.\ D {\bf 81}, 023508 (2010)
  doi:10.1103/PhysRevD.81.023508
  [arXiv:0904.4811 [gr-qc]].

\bibitem{Tsamis:1989yu} 
  N.~C.~Tsamis and R.~P.~Woodard,
  Annals Phys.\  {\bf 215}, 96 (1992).
  doi:10.1016/0003-4916(92)90301-2

\bibitem{Frob:2017apy} 
  M.~B.~Fröb,
  arXiv:1706.01891 [hep-th].

\bibitem{Miao:2010vs} 
  S.~P.~Miao, N.~C.~Tsamis and R.~P.~Woodard,
  J.\ Math.\ Phys.\  {\bf 51}, 072503 (2010)
  doi:10.1063/1.3448926
  [arXiv:1002.4037 [gr-qc]].

\bibitem{Tsamis:1992zt} 
  N.~C.~Tsamis and R.~P.~Woodard,
  Phys.\ Lett.\ B {\bf 292}, 269 (1992).
  doi:10.1016/0370-2693(92)91174-8

\end{thebibliography}
\end{document}